\documentstyle[12pt,axodraw,epsfig]{article}

\newcommand{\wt}{\widetilde}
\newcommand{\imag}{\Im {\rm m}}
\newcommand{\real}{\Re {\rm e}}
\newcommand{\tanb}{\tan \! \beta}
\newcommand{\cotb}{\cot \! \beta}
\newcommand{\mto}{m^2_{\tilde{t}_1}}
\newcommand{\mtt}{m^2_{\tilde{t}_2}}
\newcommand{\mbo}{m^2_{\tilde{b}_1}}
\newcommand{\mbt}{m^2_{\tilde{b}_2}}
\newcommand{\ghat}{\hat{g}^2}
\newcommand{\htop}{\left| h_t \right|^2}
\newcommand{\hb}{\left| h_b \right|^2}

\begin{document}

\mbox{ } \\[-1cm]
\mbox{ }\hfill KIAS--P00021\\
\mbox{ }\hfill OCHA--PP--156\\
\mbox{ }\hfill hep--ph/0005118\\
\mbox{ }\hfill \today\\

\begin{center}
{\Large\bf Probing the MSSM Higgs Boson Sector with Explicit \\[0.3mm] 
           CP Violation through Third Generation Fermion Pair\\[1mm]
	   Production at Muon Colliders}\\[1cm]
Eri Asakawa$^1$, S.Y. Choi$^2$ and Jae Sik Lee$^3$
\end{center}

\bigskip 

\begin{center}
$^1${\it Graduate School of Humanities and Sciences, Ochanomizu 
          University, \\ 1-1 Otsuka 2-chome, Bunkyo, Tokyo 112-8610, Japan} 
	  \\[1mm]
$^2${\it Department of Physics, Chonbuk National University, Chonju 561--756, 
    Korea} \\[1mm]
$^3${\it Korea Institute for Advanced Study, Seoul 130--012, Korea}
\end{center}

\bigskip
\bigskip 

\begin{abstract}
We perform a systematic study of the production of a third--generation
fermion--pair, $\mu^+\mu^-\rightarrow f\bar{f}$ for $f=\tau^-,b$, and 
$t$ in the minimal supersymmetric standard model (MSSM) with explicit 
CP violation, which is induced radiatively by soft trilinear interactions
related to squarks of the third generation.
We classify all the observables for probing the CP property of the Higgs 
bosons constructed by the initial muon beam polarization along with 
the unpolarized final fermions and with the final--fermion polarization
configuration of equal helicity, respectively. 
The observables allow for complete 
determination of CP property of the neutral Higgs bosons.
The interference between the Higgs boson and gauge boson contributions 
also could provide a powerful method for the determination of the CP 
property of two heavy Higgs bosons in the top--quark pair production near 
the energy region of the Higgs--boson resonances. 
For the lightest Higgs--boson mass there is no sizable interference between 
scalar and vector contributions for the determination of the CP property 
of the lightest Higgs boson.
We give a detailed numerical analysis to show how the radiatively--induced CP
violation in the Higgs sector of the MSSM can be measured.
\end{abstract}
\hspace{1.0 cm}PACS number(s): 12.60.Jv, 11.30.Er, 13.10.+q, 14.80.Cp

\newpage

\section{Introduction}
\label{sec:introduction}

Search for the Higgs bosons and the precise measurements on their 
properties, such as the masses, the decay widths and the decay 
branching ratios, are the most important subjects to study the 
mechanism of the electroweak symmetry breaking \cite{Hunter}.  
In the Standard Model (SM), only one physical neutral Higgs boson appears.  
On the other hand, models with multiple Higgs doublets have 
CP-even and odd neutral Higgs bosons as well as charged Higgs bosons, 
if CP is a good symmetry. Otherwise, the neutral Higgs bosons
do not have to carry any definite CP--parity.\\ 

CP violation was observed in the neutral kaon system \cite{Kaon}
and the violation in B--meson decays is strongly suggested by recent 
experiments \cite{B-meson}.
In addition, CP violation constitutes one of the crucial ingredients for
an efficient non--SM generation of the cosmological baryon asymmetry at the 
electroweak scale \cite{EWBG}. An appealing scheme of CP violation
beyond the SM can be
provided by models with an extended Higgs sector where the CP asymmetry
is broken by the ground state of the Higgs potential.
It has recently been realized that, even though the Higgs potential of
the minimal supersymmetric SM (MSSM) is CP--invariant at the tree level,
explicit CP violation in the mass matrices of the third generation squarks 
can induce sizable 
CP violation in the MSSM Higgs sector through loop corrections 
\cite{EXCP1,EXCP2,EXCP3,CDL,EXCP4}. The
CP violating phases for the third generation sfermions can be quite large,
since they contribute to the electric dipole moments (EDM's) of the electron and
neutron only at the two--loop level with no generation mixing 
\footnote{The CP--violating phases associated with the
sfermions of the first and second generations can be severely constrained by
the EDM constraints. However, there have been several suggestions
\cite{XEDM1,XEDM2,XEDM3}
to evade these constraints without suppressing the CP--violating phases.} 
in the sfermion sector \cite{CKP}.
As a result, although a one--loop effect, the 
induced CP violation in the MSSM Higgs sector can be large enough
to significantly affect Higgs phenomenology at present and future 
colliders \cite{EXCP1,EXCP2,EXCP_FC,SL1,SL2,DM}. 
In this light, it is very important to examine the possibility 
of observing these Higgs bosons and investigating their properties 
in detail.\\

A muon collider is one of the ideal machines to look for the Higgs bosons
and their properties, where $\mu^+\mu^-$ pairs can be directly converted 
into neutral Higgs--boson resonances \cite{MUCOL,S_H}. 
Polarized muons can be used to detect the Higgs bosons and
measure their properties such as masses, total widths, decay branching 
fractions and CP parities \cite{SONI,GGP}.
In this paper, we present a general formalism and a detailed analysis for
the Higgs--boson effects on the polarization observables of the
production process $\mu^+\mu^-\rightarrow f\bar{f}$ $(f=\tau^-,b$ and $t$)
using the initial muon beam polarization along with the unpolarized
final fermions and  with the final--fermion polarization configuration of
equal helicity, respectively. 
It has been recently pointed out that the amplitudes for
two Higgs states, a scalar and a pseudoscalar, can interfere with 
each other sizably if the helicities of the initial and final particles
are properly fixed and if the mass difference of these Higgs bosons
is at most of the same order as their decay widths \cite{ERI}. 
We extend the work comprehensively in order to consider Higgs bosons of 
no definite CP--parity in the MSSM with explicit CP violation induced from
the third generation squark sectors.\\

The remainder of this article is organized as follows. 
Section~\ref{sec:explicit CP violation} is devoted to 
a brief review of the explicit CP violation in the MSSM Higgs sector based
on the work \cite{CDL}. In section~\ref{sec:production} we
present the helicity amplitudes of the production of a fermion pair
in $\mu^+\mu^-$ collisions and give a comprehensive classification
of the observables according to the CP and CP$\tilde{\rm T}$ 
transformation properties that are constructed by the initial muon
beam polarization along with the final fermions of no polarization
and of equal helicity, respectively. 
In section~\ref{sec:result} we make a detailed numerical 
analysis based on a given parameter set so as to get a concrete 
estimate of the usefulness of the observables. Section~\ref{sec:conclusion}
is devoted to a brief summary of our findings and to a conclusion.

\section{Explicit CP Violation in the MSSM Higgs Sector}
\label{sec:explicit CP violation}

The existence of the non--trivial CP--violating phases in the
MSSM is due to the breakdown of supersymmetry so that the CP--violating 
phases appear in the soft--breaking parameters and the mixing among 
sparticles due to the electroweak gauge symmetry breaking. 
There are three well--known sources of CP violation in the MSSM \cite{DS}. 
The first is related to the two Higgs--boson doublets present in the model 
since both the $\mu$ parameter in the superpotential and the soft 
breaking parameter $m^2_{12}$ can be complex. Secondly, there are three more 
phases of the complex masses of the U(1)$_Y$, SU(2)$_L$ and
SU(3)$_C$ gauginos of the SM gauge group. Thirdly, the other 
CP--violating phases originate from the flavor sector of the MSSM
Lagrangian, either in the scalar soft mass matrices or the trilinear 
matrices.\\

The CP--violating phases associated with 
the off-diagonal terms of the trilinear matrices are, however,   
strongly suppressed by the same mechanism required to suppress 
the flavor changing neutral current effects. Therefore, all
these flavor--changing CP--violating phases are neglected 
in the present work such that the scalar soft mass matrices and trilinear 
parameters are flavor diagonal and the complex trilinear terms  
are proportional to the corresponding fermion Yukawa couplings.
Clearly, the Yukawa interactions of the third--generation quarks and 
squarks play the most significant role in radiative corrections to the 
Higgs sector. In this section, we give a brief review of the calculation 
\cite{CDL} of the Higgs--boson mass matrix based on the full one--loop 
effective potential, valid for all values of the relevant third--generation 
soft--breaking parameters.\\

The MSSM contains two Higgs doublets $H_1, \ H_2$, with hypercharges
$Y(H_1) = -Y(H_2) = -1/2$. Here we are only interested in the neutral
components, which we write as
\begin{eqnarray}
\label{e1}
H_1^0 = \frac{1} {\sqrt{2}} \left( \phi_1 + i a_1 \right); \ \ \ \ \
H_2^0 = \frac {{\rm e}^{i \xi}} {\sqrt{2}} \left( \phi_2 + i a_2 \right),
\end{eqnarray}
where $\phi_{1,2}$ and $a_{1,2}$ are real fields. The constant phase
$\xi$ can be set to zero at tree level, but will in general become
non--zero once loop corrections are included. \\

The mass matrix of the neutral Higgs bosons can be computed from the
effective potential \cite{CW}
\begin{eqnarray}
\label{e2}
V_{\rm Higgs}\hskip -0.3cm 
   &= \, \frac{1}{2}m_1^2\left(\phi_1^2+a_1^2\right)
     +\frac{1}{2}m_2^2\left(\phi_2^2+a_2^2\right)
     -\left|m_{12}^2\right|\left(\phi_1\phi_2-a_1 a_2\right) 
           \cos (\xi + \theta_{12}) \nonumber \\ 
   & -\left|m^2_{12}\right|
      \left(\phi_1 a_2 +\phi_2 a_1\right)\sin(\xi+\theta_{12})
     +\frac{\hat{g}^2}{8} {\cal D}^2 
     +\frac{1}{64\,\pi^2} {\rm Str} \left[
           {\cal M}^4 \left(\log\frac{{\cal M}^2}{Q^2} 
	                  - \frac{3}{2}\right)\right],
\end{eqnarray}
where we have allowed the soft breaking parameter $m^2_{12} = \left|
m^2_{12} \right| {\rm e}^{i \theta_{12}}$ to be complex, and we have
introduced the quantities
\begin{eqnarray}
\label{e5}
{\cal D} = \phi_2^2 + a_2^2 - \phi_1^2 - a_1^2; \ \ \ \ \ \hat{g}^2 
         = \frac{g^2 + g'^2}{4},
\end{eqnarray}
where the symbols $g$ and $g'$ stand for the SU(2)$_L$ and U(1)$_Y$
gauge couplings, respectively. $Q$ in Eq.~(\ref{e2}) is the
renormalization scale; the parameters of the tree--level potential, in
particular the mass parameters $m_1^2, \ m_2^2 $ and $m_{12}^2$, are
running parameters, taken at scale $Q$. The potential (\ref{e2}) is
then independent of $Q$, up to two--loop corrections.\\

The matrix ${\cal M}$ is the field--dependent mass matrix of all modes that
couple to the Higgs bosons. The by far dominant contributions come
from the third generation quarks and squarks. The (real) masses of the
former are given by
\begin{eqnarray}
\label{e3}
m_b^2 = \frac{1}{2} |h_b|^2 \left( \phi_1^2 + a_1^2
\right); \ \ \ \
m_t^2 = \frac{1}{2} |h_t|^2 \left( \phi_2^2 + a_2^2
\right), 
\end{eqnarray}
where $h_b$ and $h_t$ are the bottom and top Yukawa couplings. The
corresponding squark mass matrices can be written as
\begin{eqnarray}
{\cal M}_{\tilde t}^2 &= \mbox{$ \left( \begin{array}{cc} 
m^2_{\wt Q} + m_t^2 - \frac{1}{8} \left( g^2 - \frac{g'^2}{3} \right) {\cal D}
&
- h_t^* \left[ A_t^* \left(H_2^0 \right)^* + \mu H_1^0 \right] \\
- h_t \left[ A_t H^0_2 + \mu^* \left( H_1^0 \right)^* \right] &
m^2_{\wt U} + m_t^2 - \frac{g'^2}{6} {\cal D}
\end{array} \right); $} \nonumber\\
{\cal M}_{\tilde b}^2 &= \mbox{$ \left( \begin{array}{cc} 
m^2_{\wt Q} + m_b^2 + \frac{1}{8} \left( g^2 + \frac{g'^2}{3} \right) {\cal D}
&
- h_b^* \left[ A_b^*  \left( H_1^0 \right)^* + \mu H_2^0 \right] \\
- h_b \left[ A_b H_1^0 + \mu^* \left( H_2^0 \right)^* \right] &
m^2_{\wt D} + m_b^2 + \frac{g'^2}{12} {\cal D}
\end{array} \right). $} 
\label{e4}
\end{eqnarray}
Here, $H_1^0$ and $H_2^0$ are given by Eq.~(\ref{e1}) while $m_t^2$
and $m_b^2$ are as in Eq.~(\ref{e3}) and ${\cal D}$ has been defined
in Eq.~(\ref{e5}). In Eq.~(\ref{e4}) $m^2_{\wt Q}, \ m^2_{\wt U}$ and
$m^2_{\wt D}$ are real soft breaking parameters, $A_b$ and $A_t$ are
complex soft breaking parameters, and $\mu$ is the complex
supersymmetric Higgs(ino) mass parameter.  \\

The calculation proceeds by plugging the field--dependent 
top/bottom--(s)quark mass eigenvalues into the potential (\ref{e2}).
The mass matrix of the Higgs bosons (at vanishing external momentum)
is then given by the matrix of
second derivatives of this potential, computed at its minimum. In
order to make sure that we are indeed in the minimum of the potential,
we solve the stationarity relations, i.e. set the first derivatives of
the potential to zero. This allows us to, e.g., express $m_1^2, \
m_2^2$ and $m_{12}^2 \sin (\xi +\theta_{12})$ as functions of the
vacuum expectation values (vevs) and the remaining parameters
appearing in the loop--corrected Higgs potential. The
equations $\partial V_{\rm Higgs} / \partial a_1 = 0$ and $\partial
V_{\rm Higgs} / \partial a_2 = 0$ are linearly dependent, i.e. lead to
only one constraint on parameters for $\langle a_1
\rangle = \langle a_2 \rangle = 0$; the remaining vevs are defined
through $\langle \phi_1 \rangle^2 + \langle \phi_2 \rangle^2 = v^2
\simeq (246 \ {\rm GeV})^2$ and $\langle \phi_2 \rangle/
{\langle \phi_1 \rangle} = \tanb$.
The re--phasing invariant quantity $|m^2_{12}|\,\sin(\xi + \theta_{12})$ 
is then determined by
$|m^2_{12}|$, $m^2_{\tilde{t}_i}$ and $m^2_{\tilde{b}_i}$ as well as
by the re--phasing invariant quantities
\begin{eqnarray}
\label{e9}
\Delta_{\tilde t} = \frac { \imag(A_t \mu {\rm e}^{i \xi}) }
{\mtt - \mto} ; \ \
\Delta_{\tilde b} = \frac { \imag(A_b \mu {\rm e}^{i \xi}) }
{\mbt - \mbo} ,
\end{eqnarray}
which describe the amount of CP violation in the squark mass
matrices.  \\

The mass matrix of the neutral Higgs bosons can now be computed from
the matrix of second derivatives of the potential (\ref{e2}), where
(after taking the derivatives) $m_1^2, \ m_2^2$ and $m_{12}^2 \sin(
\xi + \theta_{12})$ are determined by the stationarity conditions.
The massless state $G^0 = a_1 \cos \beta - a_2 \sin \beta$ is
the would--be Goldstone mode ``eaten'' by the longitudinal $Z$ boson. 
We are thus left with a
squared mass matrix ${\cal M}_H^2$ for the three states $a = a_1 \sin
\beta + a_2 \cos \beta, \ \phi_1$ and $\phi_2$. This matrix is real
and symmetric, i.e. it has 6 independent entries. 
The diagonal entry for $a$ reads:
\begin{eqnarray}
\label{e10}
\left. {\cal M}^2_{H} \right|_{aa} = m_A^2 + \frac {3} {8 \pi^2}
\left\{ \frac { |h_t|^2 m_t^2 } { \sin^2 \beta} g(\mto,
\mtt) \Delta_{\tilde t}^2 + 
\frac {|h_b|^2 m_b^2 } { \cos^2 \beta} g(\mbo,
\mbt) \Delta_{\tilde b}^2 \right\},
\end{eqnarray}
and the CP--violating entries of the mass matrix,
which mix $a$ with $\phi_1$ and $\phi_2$ read:
\begin{eqnarray}
\label{e13} 
\left. {\cal M}^2_H \right|_{a \phi_1} 
   &=& \frac {3} {16 \pi^2} \left\{
       \frac { m_t^2 \Delta_{\tilde t} } {\sin \beta} \left[ g(\mto, \mtt)
       \left( X_t \cotb - 2 \htop R_t \right) 
            - \ghat \cotb \log \frac{\mtt}{\mto} \right] \right. \\  
   && \left. \hskip 1cm
     +\frac {m_b^2 \Delta_{\tilde b}} {\cos \beta} \left[ -g(\mbo,\mbt)
      \left( X_b + 2 \hb R_b' \right) + \left( \ghat - 2 \hb \right) \log
      \frac {\mbt} {\mbo} \right] \right\}, \nonumber\\
\left. {\cal M}^2_H \right|_{a \phi_2} 
   &=& \frac {3} {16 \pi^2} \left\{
       \frac {m_t^2 \Delta_{\tilde t}} {\sin \beta} \left[ -g(\mto,\mtt)
       \left( X_t + 2 \htop R_t' \right) 
            + \left(\ghat-2\htop\right)\log\frac{\mtt}{\mto} \right]\right. \\ 
   && \left. \hskip 1cm
     +\frac { m_b^2 \Delta_{\tilde b} } {\cos \beta} \left[ g(\mbo, \mbt)
      \left( X_b \tanb - 2 \hb R_b \right) 
            - \ghat\tanb\log \frac{\mbt}{\mbo} \right]\right\}. \nonumber
\end{eqnarray}
where $\Delta_{\tilde t}$ and $\Delta_{\tilde b}$ are as in
Eq.~(\ref{e9}) and the function $g(m^2_1,m^2_2)$ is given by 
\begin{eqnarray}
\label{e11}
g(m_1^2, m_2^2) = 2 - \frac {m_1^2 + m_2^2} {m_1^2 - m_2^2} \log \frac
{m_1^2} {m_2^2} .
\end{eqnarray}
The definition of the mass squared $m^2_A$ and the dimensionless quantities 
$X_{t,b}$, $R_{t,b}$ and 
$R^\prime_{t,b}$ as well as the other CP--preserving entries of 
the mass matrix squared ${\cal M}^2_{H}$  can be found 
in Ref.~\cite{CDL}.
As noted earlier, the size of these CP--violating entries is controlled by
$\Delta_{\tilde t}$ and $\Delta_{\tilde b}$. \\

The real, symmetric matrix ${\cal M}^2_H$ can be
diagonalized with an orthogonal rotation $O$;
\begin{eqnarray}
O^T\,{\cal M}^2_H\, O={\rm diag}(m^2_{H_1},m^2_{H_2},m^2_{H_3}),
\end{eqnarray}
with the ordering of $m_{H_1}\leq m_{H_2}\leq m_{H_3}$ taken
as a convention in the present work. Note that the loop--corrected
neutral Higgs--boson sector is determined by fixing the values of  
various parameters; $m_A$, $\mu$, $A_t$, $A_b$, a renormalization scale 
$Q$, $\tan\beta$, 
and the soft--breaking third generation sfermion masses, 
$m_{\tilde Q}$, $m_{\tilde U}$, 
and $m_{\tilde D}$.  The radiatively induced phase $\xi$ is no more 
an independent parameter and it can be absorbed into the 
definition of the $\mu$ parameter so that the physically meaningful
CP phases in the Higgs sector are the phases of the re--phasing 
invariant combinations $A_t \mu {\rm e}^{i\xi}$ and 
$A_b \mu {\rm e}^{i\xi}$. This neutral Higgs--boson mixing changes the 
couplings of the Higgs fields to fermions, gauge bosons, and Higgs fields 
themselves so that the effects of CP violation in the Higgs sector
can be probed through various processes \cite{EXCP_FC,SL1,SL2}.\\

\section{Third Generation Fermion Production}
\label{sec:production}

\subsection{Production amplitudes}
\label{subsec:amplitude}

The couplings of the $\gamma$ and the neutral gauge boson $Z$ to
fermions in the MSSM is described by the same interaction Lagrangian 
as in the SM:
\begin{eqnarray}
{\cal L}_{V\mu\mu}=- eQ_f \bar{f}\gamma_\mu f\,A^\mu-\frac{e}{s_Wc_W}
                    \bar{f}\left[(T_{f3}-Q_f\,s^2_W)P_-
                                   -Q_f\,s^2_WP_+ \right]
                    \gamma_\mu f\, Z^\mu\,,
\end{eqnarray}
with the chirality projection operators $P_\pm=(1\pm \gamma_5)/2$,
and those of the neutral Higgs--boson fields with 
leptons and quarks are described by the interaction Lagrangians
\begin{eqnarray}
{\cal L}_{Hff}&=& -\frac{h_l}{\sqrt{2}}\sum_{k=1}^3 
    \bar{\ell}\left[O_{2k}-is_\beta\,O_{1k}\gamma_5\right]\ell H_k
    -\frac{h_d}{\sqrt{2}}\sum_{k=1}^3 
    \bar{d}\left[O_{2k}-is_\beta\,O_{1k}\gamma_5\right] d H_k\nonumber\\
    &&  -\frac{h_u}{\sqrt{2}}\sum_{k=1}^3
        \bar{u}\left[O_{3k}- ic_\beta\,O_{1k}\gamma_5\right]u H_k\,,
\end{eqnarray}
where $h_l$, $h_d$, and $h_u$ are the lepton and quark
Yukawa couplings:
\begin{eqnarray}
h_l=\frac{g\,m_l}{\sqrt{2}m_W c_\beta};\quad
h_d=\frac{g\,m_d}{\sqrt{2}m_W c_\beta};\quad
h_u=\frac{g\,m_u}{\sqrt{2}m_W s_\beta}\,,
\end{eqnarray}
respectively. It is then clear that all the neutral Higgs bosons couple
dominantly to the third generation fermions $t$, $b$ and $\tau$
and they couple to a muon  about 200 times more strongly than to an
electron -- the {\it primary reason} for having a muon collider.\\

\vspace*{0cm} 
\begin{center}
\begin{picture}(300,100)(0,0)

\Text(5,85)[]{$\mu^-$}
\ArrowLine(10,75)(35,50)
\ArrowLine(35,50)(10,25)
\Text(5,15)[]{$\mu^+$}

\DashLine(35,50)(75,50){4}
\Text(55,37)[]{$H_k$}

\ArrowLine(75,50)(100,75)
\Text(105,85)[]{$f$}
\ArrowLine(100,25)(75,50)
\Text(105,15)[]{$\bar{f}$}

\Text(185,85)[]{$\mu^-$}
\ArrowLine(190,75)(215,50)
\Text(185,15)[]{$\mu^+$}
\ArrowLine(215,50)(190,25)

\Photon(215,50)(255,50){4}{8}
\Text(235,37)[]{$\gamma, Z$}

\ArrowLine(255,50)(280,75)
\Text(287,85)[]{$f$}
\ArrowLine(280,25)(255,50)
\Text(287,15)[]{$\bar{f}$}

\end{picture}\\
\end{center}
\smallskip
\noindent
{\bf Figure~1}: {\it The mechanisms contributing to the process
            $\mu^+\mu^-\rightarrow f\bar{f}$; three spin--0 
	    neutral--Higgs--boson exchanges
	    and spin--1 $\gamma$ and $Z$ exchanges. Here, the index $k$
	    is 1,2 or 3.  }
\bigskip 
\bigskip

The fermion pair production process $\mu^+\mu^-\rightarrow f\bar{f}$ is 
generated by five $s$--channel mechanisms: spin--1 $\gamma$ and $Z$ 
exchanges and
three spin--0 neutral--Higgs--boson exchanges, cf. Fig.~1. 
The transition matrix element of the process 
%
\begin{eqnarray}
{\cal M} = \frac{S_{ab}}{s}\left[\bar{v}(\mu^+) P_a u(\mu^-)\right]
                   \left[\bar{u}(f) P_b v(\bar{f})\right]
   +\frac{V_{ab}}{s}\left[\bar{v}(\mu^+) \gamma_\mu P_a u(\mu^-)\right]
                   \left[\bar{u}(f)\gamma^\mu P_b v(\bar{f})\right]
    \,,
\label{eq:ff production amplitude}
\end{eqnarray}
can be expressed in terms of the scalar and vector bilinear charges,
$S_{ab}$ and  $V_{ab}$, classified according to the 
chiralities $a, b = \pm $ for the right--/left--handed 
chiralities of the associated muon and produced fermion currents,
respectively. Here, the scalar bilinear charges are given by: 
\begin{eqnarray}
S_{ab}=-\frac{h_\mu h_f}{2}\sum_{k=1}^3 D_{H_k}(s)
        \left[O_{2k}-ia s_\beta O_{1k}\right] 
        \left[S^f_k -ib\, \xi_\beta\, O_{1k}\right]
       -\frac{e^2m_\mu m_f}{4m^2_W s^2_W}\, D_Z(s)\, T_{f3}\, ab\,,
\end{eqnarray}
where $h_\mu$ and $h_f$ are the initial muon and final fermion Yukawa couplings, 
respectively, and
\begin{eqnarray}
&& D_{H_k}(s)=\frac{s}{s-m^2_{H_k}+im_{H_k}\Gamma_{H_k}}\,,\qquad 
   D_Z(s)=\frac{s}{s-m^2_Z+im_Z\Gamma_Z}\,,\nonumber\\[2mm]
&& S^f_k = \left\{\begin{array}{l}
               O_{2k} \\  
               O_{3k}   
	       \end{array}\right. , \qquad\quad
   \xi_\beta = \left\{\begin{array}{ll}
               s_\beta   & {\rm for}\ \ f=l\,\,{\rm or}\,\,d \\
               c_\beta   & {\rm for}\ \ f=u\,
	       \end{array}\right. \,.
\end{eqnarray}
On the other hand, the vector bilinear charges $V_{ab}$ have
the contributions only from the $\gamma$ and $Z$ boson exchanges:
\begin{eqnarray}
V_{ab}=e^2\left[-Q_f + g^\mu_a g^f_b D_Z(s)\right]\,,
\end{eqnarray}
where $Q_f$ is the electric charge of the fermion, $g^f_\pm$ denote 
the right/left--handed couplings of the $Z$ boson to the fermions,
respectively:
\begin{eqnarray}
g^f_+ =-Q_f\tan\theta_W,\quad g^f_-=-Q_f\tan\theta_W+\frac{T_{3f}}{s_Wc_W}. 
\end{eqnarray}
The vector bilinear charges $V_{ab}$ are real in the approximation
of neglecting the $Z$ boson width $\Gamma_Z$, which is valid for
the c.m. energy away from the $Z$ boson pole.\\ 

\bigskip

\vspace*{0cm} 
\begin{center}
\begin{picture}(300,100)(0,0)

\Line(15,10)(55,90)
\Line(55,90)(285,90)
\Line(285,90)(245,10)
\Line(245,10)(15,10)

\Text(-3,50)[]{$\mu^-$}
\Line(15,50)(145,50)
\Line(145,50)(140,54)
\Line(145,50)(140,47)

\Text(298,50)[]{$\mu^+$}
\Line(155,50)(285,50)
\Line(155,50)(160,54)
\Line(155,50)(160,47)

\Vertex(150,50){3}

\Text(209,109)[]{$f$}
\Line(150,50)(200,100)
\Line(200,100)(200,95)
\Line(200,100)(197,100)

\Text(93,-7)[]{$\bar{f}$}
\Line(150,50)(100,0)
\Line(100,0)(105,0)
\Line(100,0)(100,5)

\Text(90,65)[]{$\alpha$}
\ArrowArc(75,50)(20,60,90)
\Text(232,80)[]{$\bar{\alpha}$}
\ArrowArc(235,50)(20,60,135)

\Text(62,79)[]{$P_{_T}$}
\DashLine(75,50)(95,90){3}

\Text(195,79)[]{$\bar{P}_{_T}$}
\DashLine(235,50)(255,90){3}

\Text(190,62)[]{$\Theta$}
\ArrowArc(150,50)(30,0,45)

\SetWidth{2}

\Line(75,50)(75,80)
\Line(75,80)(71,77)
\Line(75,80)(79,77)

\Line(235,50)(205,80)
\Line(205,80)(210,80)
\Line(205,80)(205,75)

\end{picture}\\
\end{center}
\bigskip
\noindent
{\bf Figure~2}: {\it The schematic description of the production plane with 
            the scattering angles $\Theta$ and $\Phi$ as well as the 
	    transverse polarization vectors $P_T$ and $\bar{P}_T$
	    with the azimuthal angles $\alpha$ and $\bar{\alpha}$
	    with respect to the scattering angle, respectively.}
\bigskip
\bigskip

Defining the polar angle of the flight direction of the fermion $f$ with 
respect to the $\mu^-$ beam direction by $\Theta$ (See Fig.~2), 
the explicit form of 
the production amplitude (\ref{eq:ff production amplitude}) can be
evaluated in the helicity basis by the 2--component spinor 
technique of Ref.~\cite{HZ}.
Denoting the $\mu^-(\mu^+)$ helicity by the first (second) index, 
the $f$ and $\bar{f}$ helicities by the remaining two indices, then 
the general form of the helicity amplitude $\langle \sigma\bar{\sigma};
\lambda\bar{\lambda}\rangle$, consisting of a scalar helicity amplitude and 
a vector helicity amplitude, reads 
\begin{eqnarray}
\langle\sigma\bar{\sigma};\lambda\bar{\lambda}\rangle 
  =\langle \sigma\bar{\sigma};\lambda\bar{\lambda}\rangle_{_S}
  +\langle \sigma\bar{\sigma};\lambda\bar{\lambda}\rangle_{_V}\,,
\end{eqnarray}
where the scalar helicity amplitude 
$\langle \sigma\bar{\sigma};\lambda\bar{\lambda}\rangle_{_S}$
is given by
\begin{eqnarray}
\langle \sigma\bar{\sigma};\lambda\bar{\lambda}\rangle_{_S}
 = -\frac{1}{4}\sum_{ab} S_{ab} 
    (a+\sigma\beta_\mu) (b-\lambda\beta)\, \delta_{\sigma\bar{\sigma}}
    \delta_{\lambda\bar{\lambda}}\,,
\label{eq:scalar amplitude}
\end{eqnarray}
and the vector helicity amplitude 
$\langle \sigma\bar{\sigma};\lambda\bar{\lambda}\rangle_{_V}$ by
\begin{eqnarray}
\langle \sigma\bar{\sigma};\lambda\bar{\lambda}\rangle_{_V}
 &=&-\frac{1}{4}\sum_{ab}V_{ab}\bigg\{
  (1+a \sigma\beta_\mu)(1+b \lambda\beta)(\lambda\sigma+\cos\Theta)
    \, \delta_{\sigma,-\bar{\sigma}}\, \delta_{\lambda,-\bar{\lambda}}\nonumber\\
 &&{ }\hskip 1.9cm  -\frac{4 m_\mu m_f}{s}(ab-\sigma\lambda\cos\Theta)\,
    \, \delta_{\sigma\bar{\sigma}}\, \delta_{\lambda\bar{\lambda}}\nonumber\\
 &&{ }\hskip 1.9cm -\frac{2m_\mu}{\sqrt{s}}(1+b \lambda\beta)(\sigma\sin\Theta)\,
    \, \delta_{\sigma\bar{\sigma}}\, \delta_{\lambda,-\bar{\lambda}}\nonumber\\
 &&{ }\hskip 1.9cm  +\frac{2m_f}{\sqrt{s}}(1+a \sigma\beta_\mu)
    (\lambda\sin\Theta)\,
    \, \delta_{\sigma,-\bar{\sigma}}\, \delta_{\lambda\bar{\lambda}}\,\bigg\}\,,
\label{eq:vector amplitude}
\end{eqnarray}
respectively, with $a,b=\pm $, $\beta_\mu=\sqrt{1-4m^2_\mu/s}$ and 
$\beta=\sqrt{1-4m^2_f/s}$.
However, the muon mass $m_\mu$ ($=106$ MeV) is extremely small compared to 
the present experimental mass bound of approximately 100 GeV on the lightest 
Higgs boson so that one can safely neglect all the terms involving the
kinematical muon mass\footnote{For consistency, the chirality--flipped 
$Z$--boson contributions to the scalar 
bilinear charges $S_{ab}$, which are also proportional to the 
muon mass kinematically, should be neglected in the massless muon limit.}.
In the approximation, the scalar and vector helicity amplitudes can be written
as:\\[3mm]
\noindent
{\bf (i) Scalar helicity amplitudes:}
\begin{eqnarray}
&& \langle ++;++\rangle_{_S} 
   =\beta_+\,S_{+-}-\beta_-\,S_{++}\,,\nonumber \\
&& \langle ++;--\rangle_{_S}
   =\beta_-\,S_{+-}-\beta_+\,S_{++}\,,\nonumber \\ 
&& \langle --;++\rangle_{_S} 
   =\beta_-\,S_{-+}-\beta_+\,S_{--}\,,\nonumber \\
&& \langle --;--\rangle_{_S}
   =\beta_+\,S_{-+}-\beta_-\,S_{--}\,, 
\end{eqnarray}
where $\beta_\pm=(1\pm\beta)/2$. 
At asymptotically high energies $\beta_+\rightarrow 1$ and
$\beta_-\rightarrow 0$, and the other scalar helicity amplitudes vanish.\\[3mm]
{\bf (ii) Vector helicity amplitudes:} 
\begin{eqnarray}
&& \langle +-;++\rangle_{_V}
   =-\sqrt{\beta_+\beta_-}\left(V_{++}+V_{+-}\right)\,\sin\Theta\,,
     \nonumber\\
&& \langle +-;+-\rangle_{_V} 
   =-\left[\,\beta_+\,V_{++}+\beta_-\,V_{+-}\right]
     (1+\cos\Theta)\,,\nonumber\\
&& \langle +-;-+\rangle_{_V}
   =+\left[\,\beta_-\,V_{++}+\beta_+\,V_{+-}\right]
     (1-\cos\Theta)\,, \nonumber\\
&& \langle +-;--\rangle_{_V} 
   =+\sqrt{\beta_+\beta_-}\left(V_{++}+V_{+-}\right)\,
     \sin\Theta\,,\nonumber\\ 
&& \langle -+;++\rangle_{_V} 
   =-\sqrt{\beta_+\beta_-}\left(V_{-+}+V_{--}\right)\,
     \sin\Theta \,,\nonumber\\
&& \langle -+;+-\rangle_{_V} 
   =+\left[\,\beta_+\,V_{-+}+\beta_-\,V_{--}\right]
     (1-\cos\Theta)\,,\nonumber\\
&& \langle -+;-+\rangle_{_V} 
   =-\left[\,\beta_-\,V_{-+}+\beta_+\,V_{--}\right]
     (1+\cos\Theta)\,, \nonumber\\
&& \langle -+;--\rangle_{_V} 
   =+\sqrt{\beta_+\beta_-}\left(V_{-+}+V_{--}\right)\,
     \sin\Theta\,, 
\end{eqnarray}
and again the other vector helicity amplitudes vanish.\\

The CP transformation leads to the relation among the transition
helicity amplitudes:
\begin{eqnarray}
\langle \sigma\bar{\sigma};\lambda\bar{\lambda}\rangle 
  &\stackrel{\rm CP}{\longleftrightarrow} &
  +(-1)^{(\sigma-\bar{\sigma})/2}(-1)^{(\lambda-\bar{\lambda})/2}
  \,\langle -\bar{\sigma},-\sigma;-\bar{\lambda},-\lambda\rangle \,,
\end{eqnarray}
or equivalently for the scalar and vector helicity amplitudes:
\begin{eqnarray}
\langle \pm\pm;\lambda\bar{\lambda}\rangle_{_S} 
  &\stackrel{\rm CP}{\longleftrightarrow} &
   +(-1)^{(\lambda-\bar{\lambda})/2}
   \langle \mp\mp;-\bar{\lambda},-\lambda\rangle_{_S}\,,\nonumber\\
\langle \pm\mp;\lambda\bar{\lambda}\rangle_{_V} 
  &\stackrel{\rm CP}{\longleftrightarrow} &
   -(-1)^{(\lambda-\bar{\lambda})/2}
   \langle \pm\mp;-\bar{\lambda},-\lambda\rangle_{_V}\,.
\end{eqnarray}
Only the simultaneous presence of the scalar and pseudoscalar couplings 
can lead to CP violation as can be explicitly checked in Eqs.~(\ref{eq:scalar
amplitude}) and (\ref{eq:vector amplitude}).
One additional useful classification is provided by the so--called ``naive"
time reversal $\tilde{\rm T}$; under the CP$\tilde{\rm T}$ transformations
the helicity amplitudes are transformed as follows:
\begin{eqnarray}
\langle \pm\pm;\lambda\bar{\lambda}\rangle_{_S} 
  &\stackrel{\rm CP\tilde{\rm T}}{\longleftrightarrow} &
   +(-1)^{(\lambda-\bar{\lambda})/2}
   \langle \mp\mp;-\bar{\lambda},-\lambda\rangle^*_{_S}\,,\nonumber\\
\langle \pm\mp;\lambda\bar{\lambda}\rangle_{_V} 
  &\stackrel{\rm CP\tilde{\rm T}}{\longleftrightarrow} &
   -(-1)^{(\lambda-\bar{\lambda})/2}
   \langle \pm\mp;-\bar{\lambda},-\lambda\rangle^*_{_V}\,.
\end{eqnarray}
We note that it is crucial to have finite $Z$ or Higgs--boson 
widths for CP$\tilde{\rm T}$ violation. In this light, it is 
very useful to take into account the CP and CP$\tilde{\rm T}$ properties of
any physical observable simultaneously so as to investigate
not only CP violation itself but also the overlapping of any pair 
of three neutral 
Higgs boson resonances effectively.\\

\subsection{Polarized production cross section} 
\label{subsec:production cross section}

The matrix element squared  for general (longitudinal or transverse) 
beam polarization can be computed either using standard
trace techniques (employing general spin projection operators), or
from the helicity amplitudes by a suitable rotation \cite{HZ} from the
helicity basis to a general spin basis. In the former case,
neglecting the muon mass in the
spin projection operators we can obtain the following approximated form
for the $\mu^\mp$ projection operators
\begin{eqnarray}
\frac{1}{2}(\not\!{p}+m)(1+\gamma_5\not\!{s})\longrightarrow
 \frac{1}{2}(1+P_L\gamma_5)\not\!{p}+\frac{1}{2}\gamma_5P_T(\cos\alpha
      \not\!{n}_1+\sin\alpha\not\!{n}_2)\not\!{p}\,,\nonumber\\
\frac{1}{2}(\not\!{\bar{p}}-m)(1+\gamma_5\not\!{\bar{s}})\longrightarrow
 \frac{1}{2}(1-\bar{P}_L\gamma_5)\not\!{\bar{p}}
 +\frac{1}{2}\gamma_5\bar{P}_T(\cos\bar{\alpha}\not\!{n}_1
  +\sin\bar{\alpha}\not\!{n}_2)\not\!{\bar{p}}\,.
\label{eq:projection}
\end{eqnarray}
Equivalently in the helicity basis the polarization weighted
matrix element squared is given by 
\begin{eqnarray}
\overline{\sum}|{\cal M}|^2
 =\sum_{\sigma\sigma'\bar{\sigma}\bar{\sigma}'}
  {\cal M}_{\sigma\bar{\sigma}} {\cal M}^*_{\sigma'\bar{\sigma}'}
  \rho^-_{\sigma\sigma'}\rho^+_{\bar{\sigma}\bar{\sigma}'}
 ={\rm Tr}\left[{\cal M}\,\rho^+{\cal M}^\dagger\rho^{-T}\right]\,,
 \label{eq:trace}
\end{eqnarray}
where ${\cal M}_{\sigma\bar{\sigma}}$ ($\sigma,\bar{\sigma}=\pm$)
denotes the helicity amplitude for any given production process 
$\mu^-(\sigma)\mu^+(\bar{\sigma})\rightarrow X$ and
the $2\times 2$ matrices $\rho^\mp$ are the polarization
density matrices for the initial $\mu^\mp$ beams:
\begin{eqnarray}
\rho^-=\frac{1}{2}\left(\begin{array}{cc}
     1+P_L                  & P_T\,{\rm e}^{-i\alpha}  \\[1mm]
    P_T\,{\rm e}^{i\alpha} & 1-P_L
                   \end{array}\right)\,,\qquad
\rho^+=\frac{1}{2}\left(\begin{array}{cc}
     1+\bar{P}_L       & -\bar{P}_T\,{\rm e}^{i\bar{\alpha}}  \\[1mm]
    -\bar{P}_T\,{\rm e}^{-i\bar{\alpha}} & 1-\bar{P}_L
                   \end{array}\right)\,.
\end{eqnarray}
Here, $P_L$ and $\bar{P}_L$ are the longitudinal polarizations of the
$\mu^-$ and $\mu^+$ beams, while $P_T$ and $\bar{P}_T$ are the degrees
of transverse polarization with $\alpha$ and $\bar{\alpha}$ being 
the azimuthal angles between the transverse polarization vectors and
the momentum vector of $f$ as shown in Fig.~2.\\

Applying the projection operators (\ref{eq:projection}) or/and evaluating
the trace (\ref{eq:trace}) leads to the polarized matrix element 
squared of the form
\begin{eqnarray}
 &&\Sigma \equiv \sum_{\lambda\bar{\lambda}}
     \sum_{\sigma\sigma'\bar{\sigma}\bar{\sigma}'}
     \langle\sigma\bar{\sigma};\lambda\bar{\lambda}\rangle
     \langle\sigma'\bar{\sigma'};\lambda\bar{\lambda}\rangle^*
     \rho^-_{\sigma\sigma'} \rho^+_{\bar{\sigma}\bar{\sigma}'}
     \nonumber \\
 &&\hskip 0.5cm = \left(1-P_L\bar{P}_L\right)\, C_1
   + \left( P_L-\bar{P}_L\right)\, C_2 \nonumber \\
 &&\hskip 0.7cm + \left(1+P_L\bar{P}_L\right)\, C_3 
   + \left( P_L+\bar{P}_L\right)\, C_4\nonumber \\
 &&\hskip 0.7cm +\,(P_T\cos\alpha+\bar{P}_T\cos\bar{\alpha})\, C_5 
                +\,(P_T\sin\alpha+\bar{P}_T\sin\bar{\alpha})\, C_6 
		\nonumber\\[1mm]
 &&\hskip 0.7cm +\,(P_T\cos\alpha-\bar{P}_T\cos\bar{\alpha})\, C_7 
                +\,(P_T\sin\alpha-\bar{P}_T\sin\bar{\alpha})\, C_8 
                \nonumber \\[1mm]
 &&\hskip 0.7cm 
        +\,(P_L\bar{P}_T\cos\bar{\alpha}+\bar{P}_L P_T\cos\alpha)\,\, C_9
        \,+\,(P_L\bar{P}_T\sin\bar{\alpha}+\bar{P}_L P_T\sin\alpha)\,\, C_{10}
	    \nonumber\\[1mm]
 &&\hskip 0.7cm 
        +\,(P_L\bar{P}_T\cos\bar{\alpha}-\bar{P}_L P_T\cos\alpha)\, C_{11}
        +\,(P_L\bar{P}_T\sin\bar{\alpha}-\bar{P}_L P_T\sin\alpha)\, C_{12}
                 \nonumber \\[1mm]
 &&\hskip 0.7cm +\,P_T\bar{P}_T\left[\,\cos(\alpha+\bar{\alpha})\, C_{13}
                + \sin(\alpha+\bar{\alpha})\, C_{14}\right]\nonumber\\[1mm]
 &&\hskip 0.7cm +\,P_T\bar{P}_T\left[\,\cos(\alpha-\bar{\alpha})\, C_{15}
   + \sin(\alpha-\bar{\alpha})\, C_{16}\right],
 \label{eq:polarized cross section}
\end{eqnarray}
where 
the coefficients $C_n$ ($n = 1$ - $16$) are defined in terms
of the helicity amplitudes by
{\small 
\begin{eqnarray*}
&& C_1=\frac{1}{4}\sum_{\lambda,\bar{\lambda}=\pm}
          \left[|\langle +-;\lambda\bar{\lambda}\rangle|^2 
              + |\langle -+;\lambda\bar{\lambda}\rangle|^2\right], \ \
   C_2=\frac{1}{4}\sum_{\lambda,\bar{\lambda}=\pm}
          \left[|\langle +-;\lambda\bar{\lambda}\rangle|^2
              - |\langle -+;\lambda\bar{\lambda}\rangle|^2\right], \nonumber\\
&& C_3=\frac{1}{4}\sum_{\lambda,\bar{\lambda}=\pm}
          \left[|\langle ++;\lambda\bar{\lambda}\rangle|^2
              + |\langle --;\lambda\bar{\lambda}\rangle|^2\right], \ \
   C_4=\frac{1}{4}\sum_{\lambda,\bar{\lambda}=\pm}
          \left[|\langle ++;\lambda\bar{\lambda}\rangle|^2
              - |\langle --;\lambda\bar{\lambda}\rangle|^2\right], \nonumber\\
&& C_5=\frac{1}{4}\,\real\sum_{\lambda,\bar{\lambda}=\pm}
\left(\langle ++;\lambda\bar{\lambda}\rangle
    - \langle --;\lambda\bar{\lambda}\rangle\right)
\left(\langle -+;\lambda\bar{\lambda}\rangle
    - \langle +-;\lambda\bar{\lambda}\rangle\right)^*,   \nonumber\\ 
&& C_6=\frac{1}{4}\,\imag\sum_{\lambda,\bar{\lambda}=\pm}
\left(\langle ++;\lambda\bar{\lambda}\rangle
    - \langle --;\lambda\bar{\lambda}\rangle\right)
\left(\langle -+;\lambda\bar{\lambda}\rangle
    + \langle +-;\lambda\bar{\lambda}\rangle\right)^*,   \nonumber\\ 
&& C_7=\frac{1}{4}\,\real\sum_{\lambda,\bar{\lambda}=\pm}
\left(\langle ++;\lambda\bar{\lambda}\rangle
    + \langle --;\lambda\bar{\lambda}\rangle\right)
\left(\langle -+;\lambda\bar{\lambda}\rangle
    + \langle +-;\lambda\bar{\lambda}\rangle\right)^*,   \nonumber\\ 
&& C_8=\frac{1}{4}\,\imag\sum_{\lambda,\bar{\lambda}=\pm}
\left(\langle ++;\lambda\bar{\lambda}\rangle
    + \langle --;\lambda\bar{\lambda}\rangle\right)
\left(\langle -+;\lambda\bar{\lambda}\rangle
    - \langle +-;\lambda\bar{\lambda}\rangle\right)^*,   \nonumber\\ 
&& C_9=\frac{1}{4}\,\real\sum_{\lambda,\bar{\lambda}=\pm}
\left(\langle ++;\lambda\bar{\lambda}\rangle
    + \langle --;\lambda\bar{\lambda}\rangle\right)
\left(\langle -+;\lambda\bar{\lambda}\rangle
    - \langle +-;\lambda\bar{\lambda}\rangle\right)^*,   \nonumber\\ 
&& C_{10}=\frac{1}{4}\,\imag\sum_{\lambda,\bar{\lambda}=\pm}
\left(\langle ++;\lambda\bar{\lambda}\rangle
    + \langle --;\lambda\bar{\lambda}\rangle\right)
\left(\langle -+;\lambda\bar{\lambda}\rangle
    + \langle +-;\lambda\bar{\lambda}\rangle\right)^*,   \nonumber\\ 
&& C_{11}=-\frac{1}{4}\,\real\sum_{\lambda,\bar{\lambda}=\pm}
\left(\langle ++;\lambda\bar{\lambda}\rangle
    - \langle --;\lambda\bar{\lambda}\rangle\right)
\left(\langle -+;\lambda\bar{\lambda}\rangle
    + \langle +-;\lambda\bar{\lambda}\rangle\right)^*,   \nonumber\\ 
&& C_{12}=\frac{1}{4}\,\imag\sum_{\lambda,\bar{\lambda}=\pm}
\left(\langle ++;\lambda\bar{\lambda}\rangle
    - \langle --;\lambda\bar{\lambda}\rangle\right)
\left(\langle +-;\lambda\bar{\lambda}\rangle
    - \langle -+;\lambda\bar{\lambda}\rangle\right)^*,   \nonumber\\ 
&& C_{13}=-\frac{1}{2}\,\real\sum_{\lambda,\bar{\lambda}=\pm}
\left[\langle -+;\lambda\bar{\lambda}\rangle
      \langle +-;\lambda\bar{\lambda}\rangle^*\right],\ \
   C_{14}=\frac{1}{2}\,\imag\sum_{\lambda,\bar{\lambda}=\pm}
\left[\langle -+;\lambda\bar{\lambda}\rangle
      \langle +-;\lambda\bar{\lambda}\rangle^*\right],             \nonumber\\
&& C_{15}=-\frac{1}{2}\,\real\sum_{\lambda,\bar{\lambda}=\pm}
\left[\langle --;\lambda\bar{\lambda}\rangle
      \langle ++;\lambda\bar{\lambda}\rangle^*\right],\ \      
    C_{16}=\frac{1}{2}\,\imag\sum_{\lambda,\bar{\lambda}=\pm}
\left[\langle --;\lambda\bar{\lambda}\rangle
      \langle ++;\lambda\bar{\lambda}\rangle^*\right].      
\end{eqnarray*}
}
{ }\\[-2.33cm]
\begin{eqnarray}
{ } 
\end{eqnarray}\\
The production cross section is then given in terms of the distribution 
$\Sigma$ in Eq.~(\ref{eq:polarized cross section}) by
\begin{eqnarray}
\frac{{\rm d}\sigma}{{\rm d}\cos\Theta\, {\rm d}\Phi}
  =\frac{N_C\, \beta}{64\,\pi^2\,s}\Sigma\,,
\end{eqnarray}
\vskip 0.2cm
\noindent
where $N_C$ is the color factor of the final fermion; 3 for the top or
bottom quark and 1 for the tau lepton. The dependence of the distribution 
$\Sigma$ on the azimuthal angle $\Phi$ of the production plane
is encoded in the angles $\alpha$
and $\bar{\alpha}$. If the azimuthal angle $\Phi$ is measured with respect
to the direction of the $\mu^-$ transverse polarization vector, the 
$\Phi$ dependence can be exhibited explicitly by taking 
\begin{eqnarray}
\alpha=-\Phi,\qquad 
\bar{\alpha}=\eta -\Phi\,,
\end{eqnarray}
where $\eta$ is the {\it rotational invariant}  difference
$\bar{\alpha}-\alpha$ of the azimuthal
angles of the $\mu^+$ and $\mu^-$ transverse polarization vectors with
respect to the production plane. 
Clearly, only the six observables $\{C_1,C_2,C_3,C_4,
C_{15}, C_{16}\}$ can be measured independently of the azimuthal angle, 
but the other ten observables, in particular, the observables involving
the SV correlations, require the reconstruction of 
the production plane. \\

The CP transformation on the polarization vectors of the initial muon beams
corresponds to the simultaneous exchanges:
\begin{eqnarray}
P_L\leftrightarrow -\bar{P}_L,\quad
P_T\leftrightarrow \bar{P}_T,\quad 
\alpha \leftrightarrow \bar{\alpha}\,. 
\label{eq:polarization CP}
\end{eqnarray}
The CP relation (\ref{eq:polarization CP}) of the polarization vectors
leads to the classification that the first 3 terms and the distributions 
$\{C_5,C_6,C_{11},C_{12},C_{13},C_{14},C_{15}\}$ in Eq.~(\ref{eq:polarized 
cross section}) are CP--even while the other six distributions 
$\{C_4,C_7,C_8,C_9,C_{10},C_{16}\}$ are CP--odd. 
Among the CP--odd observables, the three observables $\{C_4,C_7,C_9\}$ are
CP$\tilde{\rm T}$--odd and the other three observables $\{C_8,C_{10},C_{16}\}$
are CP$\tilde{\rm T}$--even. Here, it will be noteworthy to emphasize again 
that one crucial requirement for having a large 
CP$\tilde{\rm T}$--odd observable is the presence of so--called CP--preserving
re-scattering phases which can be provided by the Higgs boson
propagators with relatively large widths in the present work.

\subsection{Initial spin correlations}

Let us now express the coefficients $C_i$ ($i=1$ to 16) in terms of the 
scalar and vector bilinear and quartic charges \cite{SZ} 
as well as the newly--introduced notations:\\[-1mm]
\begin{eqnarray}
&& R_1=\sum_{ab}S_{ab},\quad\,
   R_2=\sum_{ab}ab S_{ab},\quad
   R_3=\sum_{ab}a S_{ab},\quad\,
   R_4=\sum_{ab}b S_{ab},\nonumber\\
&& W_1=\sum_{ab}V_{ab},\quad
   W_2=\sum_{ab}ab V_{ab},\quad
   W_3=\sum_{ab}a V_{ab},\quad
   W_4=\sum_{ab}b V_{ab},
\end{eqnarray}
which are to be employed for the interference between the scalar and
vector contributions. It is worthwhile to note that both
$R_3$ and $R_4$ vanish in the CP--invariant theory.
The sixteen polarization distributions can be then
classified as follows:\\[3mm]
\noindent
{\bf (i) Scalar--scalar (SS) correlations:}
\begin{eqnarray}
&& C_{\,3}[++]\,=\frac{1}{2}\left[(1+\beta^2)\, S_1
                 -(1-\beta^2)\, S_2\right],\nonumber\\
&& C_{\,4}[--]\, = \frac{1}{2}\left[(1+\beta^2)\, S'_1
                 -(1-\beta^2)\, S'_2\right],\nonumber\\
&& C_{15}[++]=\frac{1}{2}\left[(1+\beta^2)\, S_4
                 - (1-\beta^2)\,S_5\right],\nonumber\\
&& C_{16}[-+]=\frac{1}{2}\left[(1+\beta^2)\, S'_4
                 - (1-\beta^2)\,S'_5\right],
\end{eqnarray}
where the first and second signatures in the square brackets are 
for the CP and CP$\tilde{\rm T}$ parities of the corresponding observable, 
respectively,  and the relevant scalar quartic charges are defined as
\begin{eqnarray}
&& S_1=\frac{1}{4}
       \left[|S_{++}|^2+|S_{--}|^2+|S_{+-}|^2+|S_{-+}|^2\right]\,,\nonumber\\
&& S'_1=\frac{1}{4}
        \left[|S_{++}|^2+|S_{+-}|^2-|S_{-+}|^2-|S_{--}|^2\right]\,,\nonumber\\
&& S_2=\frac{1}{2}\,\real
       \left[S_{++}S^*_{+-}+S_{--}S^*_{-+}\right]\,,\quad
   S'_2=\frac{1}{2}\,\real
        \left[S_{++}S^*_{+-}-S_{--}S^*_{-+}\right]\,,\nonumber\\
&& S_4=\frac{1}{2}\,\real
       \left[S_{++}S^*_{-+}+S_{--}S^*_{+-}\right]\,,\quad
   S'_4=\frac{1}{2}\,\imag
        \left[S_{++}S^*_{-+}-S_{--}S^*_{+-}\right]\,,\nonumber\\
&& S_5=\frac{1}{2}\,\real
       \left[S_{++}S^*_{--}+S_{+-}S^*_{-+}\right]\,,\quad
   S'_5=\frac{1}{2}\,\imag
        \left[S_{++}S^*_{--}+ S_{+-}S^*_{-+}\right]\,,
\end{eqnarray}\\
{\bf (ii) Vector--vector (VV) correlations:}
\begin{eqnarray}
&& C_{\,1}[++]\, = (1+\beta^2\cos^2\Theta)\, V_1+(1-\beta^2)\, V_2
               +2\beta\, V_3\cos\Theta,\nonumber\\
&& C_{\,2}[++]\, = (1+\beta^2\cos^2\Theta)\, V'_1+(1-\beta^2)\, V'_2
               +2\beta\, V'_3\cos\Theta,\nonumber\\
&& C_{13}[++]= \beta^2\sin^2\Theta\, V_4,\nonumber\\
&& C_{14}[+-]= \beta^2\sin^2\Theta\, V'_4,
\end{eqnarray}
where the quartic charges $V_3$ and $V'_3$ are given by  
\begin{eqnarray}
&& V_3 =\frac{1}{4}
       \left[|V_{++}|^2+|V_{--}|^2-|V_{+-}|^2-|V_{-+}|^2\right]\,,\nonumber\\
&& V'_3=\frac{1}{4}
       \left[|V_{++}|^2-|V_{+-}|^2+|V_{-+}|^2-|V_{--}|^2\right]\,,
\end{eqnarray}
and the other vector quartic charges are defined in the same way as
the scalar quartic charges with the notation $S$ replaced by 
$V$ everywhere. We note that the scalar as well as vector quartic charges 
defined as an imaginary part of the bilinear--charge correlations might be 
non-vanishing only when there are complex CP--violating couplings 
or/and CP--preserving phases like re-scattering phases or finite widths 
of the intermediate particles. So, if there are no CP--preserving phases, 
non--vanishing values of these quartic charges signal CP violation 
in the given process.\\[3mm]
%
%
{\bf (iii) Scalar--vector (SV) correlations:}
\begin{eqnarray}
&& C_{\,5}[++]\, =+\frac{m_f}{4\sqrt{s}}\,\beta\sin\Theta\,
        \real(W_3\,R^*_1),\quad 
   C_{\,6}[+-]\,\, =+\frac{m_f}{4\sqrt{s}}\,\beta\sin\Theta\,
        \imag(W_1\,R^*_1),\nonumber\\
&& C_{\,7}[--]\, =-\frac{m_f}{4\sqrt{s}}\,\beta\sin\Theta\,
        \real(W_1\,R_3^*), \quad
   C_{\,8}[-+]\,\, =-\frac{m_f}{4\sqrt{s}}\,\beta\sin\Theta\,
        \imag(W_3\,R_3^*), \nonumber\\
&& C_{\,9}[--]\, =+\frac{m_f}{4\sqrt{s}}\,\beta\sin\Theta\,
        \real(W_3\,R_3^*), \quad\,
   C_{10}[-+]  =+\frac{m_f}{4\sqrt{s}}\,\beta\sin\Theta\,
        \imag(W_1\,R^*_3),\nonumber\\ 
&& C_{11}[++]  =+\frac{m_f}{4\sqrt{s}}\,\beta\sin\Theta\,
        \real(W_1\,R^*_1),\quad
   C_{12}[+-]  =+\frac{m_f}{4\sqrt{s}}\,\beta\sin\Theta\,
        \imag(W_3\,R_1^*). 
\end{eqnarray}
All the SV correlations are proportional to the mass of the
final--state fermion divided by the c.m. energy so that they are
strongly suppressed at high energies. In this light, they are sizable
only in the production of the top--quark pair because of the largest
fermionic mass.

\subsection{Final spin correlations for Higgs interference effects }

Various types of experimental observables by use of the initial muon
polarization have been considered in the previous section for 
determining the CP character of the neutral MSSM Higgs bosons. 
In addition, spin correlations of $t\bar{t}$ or $\tau^+\tau^-$
in the final state can also probe the CP nature of the Higgs boson,
independently of or combined with initial beam polarization.
If the production plane or directly the momentum directions of two 
final fermions are determined, the secondary decays of the primary 
final state fermions can allow a complete analysis of their spin or 
helicity directions. However, even if the production plane is not 
reconstructed, one can obtain the fermion--pair polarization combination 
by statistically studying decay products from the correlated and polarized
$f\bar{f}$. It is then natural that 
the production distribution needs to be integrated over the azimuthal 
angle $\Phi$.\\

The $s$--channel Higgs boson (spin--0) exchange populates the 
equal $\mu^-\mu^+$ helicity combinations, i.e. $\sigma=\bar{\sigma}$.
This results in the correlations of $f\bar{f}$ polarization with  
$\lambda=\bar{\lambda}$ by angular momentum conservation, while
the SM background channels yield dominantly the $(+-)$ or
$(-+)$ polarization combination. In addition, the CP transformation
interchanges the $(++)$ and $(--)$ helicity configurations.
In the light of these two arguments combined with the one in the previous 
paragraph, it will be valuable to consider the $\Phi$--independent 
polarization observable\footnote{In principle, there can exist the  
$\gamma$ and $Z$--exchange terms corresponding to the polarization
combinations $(1-P_L \bar{P}_L)$ 
and $(P_L-\bar{P}_L)$ of the initial $\mu^+\mu^-$ beams. 
However, those terms turn out to be zero for the real vector and axial--vector
couplings in the approximation of neglecting the $Z$--boson width.}:
\begin{eqnarray}
&&\Delta \equiv
\sum_{\lambda}
\sum_{\sigma\sigma'\bar{\sigma}\bar{\sigma}'}
\lambda\, \langle\sigma\bar{\sigma};\lambda\lambda\rangle
\langle\sigma'\bar{\sigma}';\lambda\lambda\rangle^*
\rho^-_{\sigma\sigma'} \rho^+_{\bar{\sigma}\bar{\sigma}'}
\nonumber\\
 &&\hskip 0.5cm  =  \left(1+P_L\bar{P}_L\right)\, D_1 
   + \left( P_L+\bar{P}_L\right)\, D_2
 +\,P_T\bar{P}_T\left[\,\cos\eta\, D_3
   + \sin\eta\, D_4\right],
 \label{eq:difference}
\end{eqnarray}
where the distributions $D_i$ ($i=1$ to 4) are defined in terms of the
helicity amplitudes as 
\begin{eqnarray}
&& D_1[--]=\frac{1}{4}\sum_\lambda\, \lambda
          \left[|\langle ++;\lambda\lambda\rangle|^2
              + |\langle --;\lambda\lambda\rangle|^2\right], \nonumber\\
&& D_2[++]=\frac{1}{4}\sum_\lambda\, \lambda
          \left[|\langle ++;\lambda\lambda\rangle|^2
              - |\langle --;\lambda\lambda\rangle|^2\right], \nonumber\\
&& D_3[--]=-\frac{1}{2}\,\real\sum_\lambda\, \lambda
       \left[\langle --;\lambda\lambda\rangle
             \langle ++;\lambda\lambda\rangle^*\right],\nonumber\\      
&& D_4[+-]=-\frac{1}{2}\,\imag\sum_\lambda\, \lambda
       \left[\langle --;\lambda\lambda\rangle
             \langle ++;\lambda\lambda\rangle^*\right],      
\end{eqnarray}
where the first and second signatures in the square brackets are
for the CP and CP$\tilde{\rm T}$ parities of the corresponding 
observable.

\subsection{CP$\tilde{\rm T}$--even and odd combinations of Higgs--boson
            propagators}

The CP$\tilde{\rm T}$ parity of each observable plays a crucial role in
determining the interference pattern of the Higgs bosons among themselves and 
with the $\gamma$ and $Z$ bosons appearing in the  observable; 
every CP$\tilde{\rm T}$--even (odd) observable involving only the scalar
contributions depends on the real 
(imaginary) part of the combination $D_{H_k}D^*_{H_l}$ of each pair 
of two Higgs--boson propagators. The CP$\tilde{\rm T}$--odd observable
$C_{14}$ involving only the vector contributions is negligible because 
the $Z$--boson width effect is significantly small for the energy far from the
$Z$--boson pole. When the $Z$--boson width is neglected, all the vector
couplings are real. As a result, the interference terms between the
scalar and vector contributions depends on the real or imaginary parts
of the Higgs--boson propagators $D_{H_k}$ itself.\\

For the sake of discussion in the following, 
let us introduce for the real and imaginary parts the abbreviated notations:
\begin{eqnarray}
&& {\cal S}_{kl}\equiv\,\, \real [D_{H_k} D^*_{H_l}]
                \, = \frac{(s-m^2_{H_k})(s-m^2_{H_l})
		       +m_{H_k}m_{H_l}\Gamma_{H_k}\Gamma_{H_l}}{
		        [(s-m^2_{H_k})^2+m^2_{H_k}\Gamma^2_{H_k}]
		        [(s-m^2_{H_l})^2+m^2_{H_l}\Gamma^2_{H_l}]},
			\nonumber\\
&& {\cal D}_{kl}\equiv\imag [D_{H_k} D^*_{H_l}]
                = \frac{(s-m^2_{H_k})\, m_{H_l}\Gamma_{H_l}
		       -(s-m^2_{H_l})\, m_{H_k}\Gamma_{H_k}}{
		        [(s-m^2_{H_k})^2+m^2_{H_k}\Gamma^2_{H_k}]
		        [(s-m^2_{H_l})^2+m^2_{H_l}\Gamma^2_{H_l}]}.
\end{eqnarray}
It is worthwhile to note two points; (a) the denominator reveals a typical 
two--pole structure so that the Higgs--boson contributions are greatly 
enhanced at the poles; (b) the numerator of ${\cal S}_{kl}$ is negative in the 
middle of two resonances with a mass splitting larger than their typical
widths but positive otherwise, while the numerator of ${\cal D}_{kl}$ 
is always positive (negative) if 
$m_{H_l} \geq m_{H_k}$ ($m_{H_k}\geq m_{H_l}$) and linearly increasing 
(decreasing ) if $m_{H_l}\Gamma_{H_l}\geq m_{H_k}\Gamma_{H_k}$
($m_{H_k}\Gamma_{H_k}\geq m_{H_l}\Gamma_{H_l}$), respectively.\\

On the other hand, the SV interference terms depend on the real and
imaginary parts of each Higgs--boson propagators $D_{H_k}$, the form of which is
given by
\begin{eqnarray}
\real[D_{H_k}]=\frac{s-m^2_{H_K}}{(s-m^2_{H_k})^2+m^2_{H_k}\Gamma^2_{H_k}},
  \qquad
\imag[D_{H_k}]=-\frac{m_{H_K}\Gamma_{H_k}}{(s-m^2_{H_k})^2
                             +m^2_{H_k}\Gamma^2_{H_k}}.
\end{eqnarray}
Note that the real part changes its sign whenever the c.m. energy crosses
the pole, but the imaginary part is always negative.\\

Combining the coefficients from  the mixing matrix elements with 
those propagator--dependent parts enables us to make a straightforward
qualitative understanding of the $\sqrt{s}$ dependence of each observable. 
This will be demonstrated in detail in the following section with a 
concrete numerical example.

\section{Numerical Results}
\label{sec:result}
 
We are now ready to present some numerical results. It is known that 
loop--induced CP violation in the Higgs sector can only be large if
both $|\mu|$ and $|A_t|$ (or $|A_b|$, if $\tan\beta\gg 1$) are 
sizable \cite{EXCP1,EXCP2,EXCP3,CDL}. We therefore choose 
$|A_t|=|A_b|=2m_{\tilde{Q}}$. 
For definiteness we will present results only for the $t\bar{t}$ mode by taking
$\tan\beta=3,10$ when probing the region around two heavy 
Higgs--boson resonances, but we will give a qualitative description
of probing the CP property of the lightest Higgs boson. 
We take the running Yukawa couplings
by including the QCD loop effects depending on the gluino mass properly.
Since for moderate values of $\tan\beta$ the contributions from the
(s)bottom sector are still quite small, our results are not sensitive
to $m_{\tilde{D}}$ and $A_b$; we therefore fix $m_{\tilde{D}}=
m_{\tilde{U}}=m_{\tilde{Q}}$, although different values for the 
SU(2) doublet and singlet soft breaking squark masses, $m_{\tilde{Q}}\neq 
m_{\tilde{U}}$ are allowed, and also take equal phases for $A_t$ and $A_b$.
Since we are basically interested in distinguishing the CP 
non--invariant Higgs sector from the CP invariant one, we take
for the re-phasing--invariant phase $\Phi_{A\mu}$ of $A_{t,b}\mu\, 
{\rm e}^{i\xi}$ two values; 0 (CP invariant case) and $\pi/2$ 
(maximally CP violating case). On the other hand, we fix for simplicity
the other real mass parameters and couplings except for $\tan\beta$ 
as follows:
\begin{eqnarray}
&&  m_{A} = 0.4\,{\rm TeV},\ \
   |A_{t,b}|= 1.0\,{\rm TeV},\ \
   |\mu|=1.0\, {\rm TeV},\nonumber\\
&& m_{\tilde{g}}= 0.5\, {\rm TeV},\quad 
   m_{\tilde{Q}}=m_{\tilde{U}}=m_{\tilde{D}}=0.5\, {\rm TeV}\,.
\label{eq:parameter set}
\end{eqnarray}
Finally, we take for the top and bottom quark running masses 
$\bar{m}_t(m_t)= 165$ GeV and $\bar{m}_b (m_b)=4.2$ GeV.
For the given $m_A$ much larger than $m_Z$, two neutral Higgs bosons
have almost degenerate masses:
\begin{eqnarray}
\tan\beta=\,\,\, 3,\, \Phi_{A\mu}=\,0 &:& 
      m_{H_2}=400.0\,\, {\rm GeV},\ \ m_{H_3}=400.4\,\, {\rm GeV},\nonumber\\
\tan\beta=\,\, 3,\, \Phi_{A\mu}=\frac{\pi}{2} &:& 
      m_{H_2}=396.6\,\, {\rm GeV},\ \ m_{H_3}=404.5\,\, {\rm GeV},\nonumber\\
\tan\beta=10,\, \Phi_{A\mu}=\,0 &:& 
      m_{H_2}=397.5\,\, {\rm GeV},\ \ m_{H_3}=400.0\,\, {\rm GeV},\nonumber\\
\tan\beta=10, \Phi_{A\mu}=\frac{\pi}{2} &:& 
      m_{H_2}=397.3\,\, {\rm GeV},\ \ m_{H_3}=400.5\,\, {\rm GeV}.
\label{eq:higgs masses}
\end{eqnarray}
As a result, a significant overlapping, i.e. interference between two 
Higgs--boson resonances is expected. \\

In the previous section, we have listed 16 polarization observables 
constructed solely by initial muon polarizations under the assumption that
the production plane is (at least statistically) reconstructed, and
4 additional polarization observables by combining initial muon 
polarizations and final equal helicity configurations only for the
case when the production plane does not have to be explicitly reconstructed.
For some polarization observables it is also important to experimentally 
identify the electric charges of the produced fermion pair from
their decay products. Taking into account all these requirements naturally leads
us to the following classification; for the energy range around the heavy 
Higgs boson resonances the production of a top quark pair is 
recommended to be used (when the heavy Higgs--boson masses are larger
than twice the top--quark mass), and for the energy close to the lightest 
Higgs mass, the tau lepton or bottom quark pairs are recommended to be
considered\footnote{The $b\bar{b}$ decay mode of the lightest Higgs boson, 
even if it has the largest branching fraction, is not useful for certain 
observables because of the difficulty in determining the polarization
and electric charge of the produced bottom quark and
anti--quark experimentally.}\\

\subsection{Two heavy Higgs bosons for top--pair production}

In the MSSM two heavy Higgs bosons are (almost) degenerate with
a mass splitting comparable to or less than their widths 
for $M_A\gg m_Z$ as in the case of the parameter set (\ref{eq:parameter set}),
and one of them is CP--even and the other CP--odd in the CP invariant 
theory. However, the CP--violating neutral Higgs--boson mixing can cause
a mass splitting larger than their typical widths and it can lead to several
significant non--vanishing CP--odd observables, in particular,
in the $t\bar{t}$ mode. In this section, we investigate in detail
the possibility of probing those aspects through the observables classified
in the previous section by considering the cases without and
with direct reconstruction of the production plane separately.

\subsubsection{Without direct reconstruction of the production plane}

If the production plane is not directly reconstructed but only the 
helicities of the produced top quark and anti-quark are statistically
determined, all the interference between the scalar and vector 
contributions is averaged away over the azimuthal angle $\Phi$ and only 
the observables, $\{C_1, C_2, C_3, C_4, C_{15}, C_{16}; D_1, D_2, D_3, D_4\}$ 
can be reconstructed. \\

Figures~\ref{fig:fig1} and \ref{fig:fig2} exhibit six independent 
observables 
of definite CP and CP$\tilde{\rm T}$ parities near the region
of the heavy Higgs--boson resonances;
\begin{eqnarray*}
{\rm (a)}\, \sigma_{_{RL}}[++], \ \  {\rm (b)}\, \sigma_{_{LR}}[\,++], \ \ 
   {\rm (c)/(d)}\,\, \frac{\sigma_{_{RR}}\pm\sigma_{_{LL}}}{2}[\,\pm\pm], \ \ 
   {\rm (e)}\,\, \sigma_{\Vert}[++], \ \
   {\rm (f)}\,\, \sigma_{\bot}[-+],
\end{eqnarray*}
for $\tan\beta=3,10$ in Figs.~\ref{fig:fig1} and \ref{fig:fig2}, respectively. 
The observables obtained by controlling solely the 
muon and anti--muon polarizations are given by
\begin{eqnarray}
\sigma_{_{RL/LR}}  &=& \frac{N_C\, \beta}{16\,\pi\,s}\,\int d\cos\Theta
                   \,\left[\, C_1\pm C_2\,\right]\,,\nonumber\\
\sigma_{_{RR/LL}}  &=& \frac{N_C\, \beta}{8\,\pi\,s}\,
              \,\left[\, C_3\pm C_4\,\right]\,, \nonumber\\ 
\sigma_{\,\Vert/\bot}&=& \frac{N_C\, \beta}{16\,\pi\,s}\,(\pm)\,C_{15/16}\,.
\label{eq:observable1}
\end{eqnarray}
%
As shown in the frames (a) and (b) the spin--1 (LR) and (RL) observables
$\sigma_{LR}$ and $\sigma_{RL}$ in both figures are almost constant and 
the former is almost twice larger than the latter, showing the large 
left--right asymmetry.  On the other hand, the average of the observables 
$\sigma_{LL}$ and $\sigma_{RR}$ showing the Higgs--boson contributions 
are peaked on each heavy Higgs--boson resonance and the size at each pole 
is comparable to that of the spin--1 observable.  In the CP invariant case
($\Phi_{A\mu}=0$) with $\tan\beta=3$ the parameter set (\ref{eq:parameter set}) 
with $\tan\beta=3$ accidentally generates two extremely degenerate Higgs bosons
as denoted by a single resonance line in the frame (c) as well as a single 
oscillating pattern in the frame (e) for $\sigma_{\Vert}$ of
Fig.~\ref{fig:fig1} unlike the case for $\tan\beta=10$ with a finite mass 
splitting of about 3 GeV. The CP--odd observables $(\sigma_{RR}-\sigma_{LL})/2$ 
and $\sigma_{\bot}$ are identically zero for both $\tan\beta=3$ and $10$
as they ought to be.  On the other hand, in the CP non--invariant case 
a large mass splitting between two Higgs--boson resonances is developed 
for $\tan\beta=3$, implying a large CP--violating mixing between two Higgs 
bosons, while the splitting in the case of $\tan\beta=10$ is not so much
enlarged. A more quantitative understanding about the mass splitting
can be obtained from Eq.~(\ref{eq:higgs
masses}). In this case, the CP--odd observables are non--vanishing. Note that 
observable $\sigma_{\bot}$ are quite sizable on the poles, especially, 
in the case of $\tan\beta=3$. \\

The observable $(\sigma_{_{RR}}-\sigma_{_{LL}})/2$ in the frame (d) is 
CP$\tilde{\rm T}$--odd, involving the propagator combination ${\cal D}_{kl}$,
but the observables $\sigma_{\Vert}$ and $\sigma_{\bot}$ in the frames (e)
and (f) are CP$\tilde{\rm T}$--even, involving the propagator combination 
${\cal S}_{kl}$.
Examining the analytic structure of those observables and taking into account
the parameter set (\ref{eq:parameter set}), we find that
(i) the coefficient related with the CP--odd (CP--even) resonance
pole is negative (positive), respectively; 
(ii) for $\tan\beta=3$ the lighter (heavier) resonance of the heavy Higgs
bosons is dominantly CP--odd (CP--even) in both the CP invariant and 
non--invariant cases, respectively;
(iii) for $\tan\beta=10$ the lighter resonance is dominantly CP--even
(CP--odd) and the heavier one CP--odd (CP--even) in the CP invariant
(non--invariant) case for $\Phi_{A\mu}=0$ ($\pi/2$), respectively.
Combining all these features leads to a symmetric resonance pattern 
in the frame (d) and anti--symmetric patterns in the frames (e) and (f)
in both figures and a remarkable sign flipping at each resonance pole
in the frame (e) of Fig.~\ref{fig:fig2}.\\

We present in Fig.~\ref{fig:fig3} three observables\footnote{The sum 
$\Delta_{_{RR}}+\Delta_{_{LL}}$ is strongly suppressed in the present case 
due to the orthogonality of the neutral--Higgs 
mixing matrix $O$: $\sin\beta$ is very close to unity even for $\tan\beta=3$ 
such that the product sum $O_{2k}O_{2l}+s^2_\beta\, O_{1k}O_{1l}$ is
approximately $\delta_{kl}$, i.e. vanishing in the present case because 
of $k\neq l$ forced by the antisymmetric combination 
$O_{1k}S^f_l-S^f_k O_{1l}$ with $S^f_l=O_{3l}$ for $f=t$.}
related with the difference between the cross sections with the
$(++)$ and $(--)$ helicity configurations:
\begin{eqnarray*}
\frac{\Delta_{_{RR}}-\Delta_{_{LL}}}{2}[++],\quad
\Delta_{\bot}[+-],\quad
\Delta_{\Vert}[--]\,,
\label{eq:observable22}
\end{eqnarray*}
near the heavy Higgs--boson resonances, the explicit form of
which is obtained by integrating the distributions $\{D_2,D_3,D_4\}$ over 
the production angles as 
\begin{eqnarray}
\Delta_{_{RR/LL}} = \frac{N_C\, \beta}{8\,\pi\,s}\,
              \,\left[\, D_1\pm D_2\,\right]\,, \quad  
\Delta_{\,\Vert/\bot}= \frac{N_C\, \beta}{16\,\pi\,s}\,D_{3/4}.
\label{eq:observable2}
\end{eqnarray}
It should be noted that every observable $D_i$ ($i=1$ to 4)
involves the antisymmetric combination $(O_{1k} O_{3l}-O_{3k} O_{1l})$ 
forcing $k\neq l$ ($k,l=1,2,3$). Because two heavy Higgs bosons 
exhibit almost a typical two--state mixing for the parameter set
(\ref{eq:parameter set}), there exist simply one 
coupling combination formed by the mixing matrix elements for each observable.
As a result, the $\sqrt{s}$ dependence of each observable is (almost) 
completely determined by that of the propagator combinations
${\cal S}_{23}$ or ${\cal D}_{23}$, depending on whether the observable
is CP$\tilde{\rm T}$--even or CP$\tilde{\rm T}$--odd. As two Higgs
bosons are extremely degenerate in the CP invariant case with $\tan\beta=3$
a single resonance peak (solid line) is shown in the first two upper frames of
Fig.~\ref{fig:fig3}. However, we note in the upper part of
Fig.~\ref{fig:fig3} that (i) in the CP non--invariant 
case ($\Phi_{A\mu}= \pi/2$) a large mass splitting is developed;
and (ii) the CP$\tilde{\rm T}$--even observable $(\Delta_{RR}-\Delta_{LL})/2$ 
changes its sign at each resonance pole as ${\cal S}_{23}$ does, while
the other two CP$\tilde{\rm T}$--odd observables have a two--pole 
structure of the same sign. On the other hand, the mass splitting between 
two heavy Higgs bosons for $\tan\beta=10$ is not so different
in the CP invariant and non--invariant cases. 
As explained in the case of $\tan\beta=3$,
one can see a sign change at each resonance pole in the CP$\tilde{\rm T}$--even
observable and a typical two--pole structure in the CP$\tilde{\rm T}$--odd
observables. It is also noteworthy that the observable $\Delta_{\bot}$ is
negative (positive) for the CP invariant (non--invariant) case.
Finally, the typical size of the CP--odd observables is very small compared 
to that of the CP--even observables.

\subsubsection{With direct reconstruction of the production plane}

All the terms involving interference between the scalar and
vector contributions requires a reasonable identification of the
production plane and they are proportional to the mass of the final 
fermion. The $\tau^+\tau^-$ mode with two final neutrinos 
escaping detection can be used at high energies where the produced 
tau leptons are very energetic and therefore their decay products
fly along the original tau direction to a very good approximation. 
Nevertheless, the small mass leads to very small interference effects.
On the other hand, the production plane in the $t\bar{t}$ mode can
be determined without big difficulty and the top--quark mass is
almost 100 times larger than the tau--lepton mass. In this light, 
we will concentrate on the $t\bar{t}$ mode which could generate
significant interference effects.\\

Among the eight interference terms the four terms 
$\{C_7,C_8,C_9,C_{10}\}$ are CP--odd and the other four terms
$\{C_5,C_6,C_{11},C_{12}\}$ are CP--even. It is noteworthy that
all the CP--odd terms are proportional to $R_3$ while all 
the CP--even observables to $R_1$; as a matter of fact the CP properties 
of the interference terms originates from those of $R_1$ and $R_3$
as can be worked out by their explicit form for the $t\bar{t}$ mode:
\begin{eqnarray}
R_1= -2h_\mu h_f\, \sum_{k=1}^3 D_{H_k}\, O_{2k} O_{3k}\,,\quad 
R_3= 2ih_\mu h_f\, s_\beta\, \sum_{k=1}^3 D_{H_k}\, O_{1k} O_{3k}\,.
\end{eqnarray}
We note that (i) $R_1$ describes the CP--preserving 
mixing between two CP--even states, $\phi_1$ and $\phi_2$, so that
the relevant observables can be used to select the dominantly
CP--even Higgs--boson states; (ii) the coefficients $O_{22}O_{32}$ and
$O_{23}O_{33}$ have an equal sign while the coefficients $O_{12}O_{32}$ 
and $O_{13}O_{33}$ have an opposite sign; (iii) $R_3$ the CP--violating mixing 
between $\phi_2$ and the CP--odd state $a$.
In addition, the coefficients $W_{1,3}$ involving vector
bilinear charges are (almost) real to a very good approximation
so that the SV interference terms depends on the real and
imaginary parts of each Higgs--boson propagator $D_{H_k}$ itself.\\

Certainly we need to apply an appropriate $\Phi$--dependent 
weight function to extract each interference term; for example,
($\cos\alpha-\cos\bar{\alpha}$) with $P_T=\bar{P}_T=1$ taken
can be used as a weight function to extract $C_7$. The extraction 
efficiency should be determined with the detailed information on various 
experimental machine parameters. 
So, we will not provide any further detailed procedure to 
extract those observables, but concentrate ourselves on the observables
$C_i$ themselves. 
Let us define the observables $\sigma_i$ ($i=5$ to $12$) as:
\begin{eqnarray}
\sigma_i = \frac{N_C\,\beta}{32\pi\,s}\, \int {\rm d}\cos\Theta\,C_i\,,
\label{eq:observable3}
\end{eqnarray}
so that $\sigma_i$ has the same CP property as $C_i$. \\

Taking the maximal CP--violating phase $\Phi_{A\mu}=\pi/2$, we present
in Fig.~\ref{fig:fig4} the CP--odd observables
$\{\sigma_7[--],\sigma_8[-+],\sigma_9[--],\sigma_{10}[-+]\}$ for the 
$t\bar{t}$ mode near the heavy Higgs--boson resonances for 
$\tan\beta=3$ (solid line) and $\tan\beta=10$ (dashed line) with 
the SUSY parameter set (\ref{eq:parameter set}). 
The CP$\tilde{\rm T}$--even observables $\sigma_8$ and $\sigma_{10}$
changes their sign at each pole as the real part of each Higgs--boson
propagator changes its sign. On the other hand, the fact that the coefficients
$O_{12}\,O_{32}$ and $O_{13}O_{33}$ have an opposite sign is responsible 
for the sign--flipping pattern at the poles, appearing in the left frames of
Fig.~\ref{fig:fig4}. On the whole, among those CP--odd observables
$\sigma_7$ and $\sigma_{10}$ are sizable while the other two observables
are small in size.\\

Figure~\ref{fig:fig5} shows the CP--even observables
$\left\{\sigma_5[++],\sigma_6[+-],\sigma_{11}[++],\sigma_{12}[+-]\right\}$ 
for the 
$t\bar{t}$ mode near the heavy Higgs--boson resonances for 
$\Phi_{A\mu}=0$ (solid line) and $\Phi_{A\mu}=\pi/2$ (dashed line) with 
the SUSY parameter set (\ref{eq:parameter set}). In this case, we take 
$\tan\beta=3$ causing a larger CP--violating Higgs--boson
mixing than $\tan\beta=10$. As explained before, the observables dependent 
on the coefficients $O_{2k}O_{3k}$ single the dominantly CP--even states 
out. As can be seen in every frame of Fig.~\ref{fig:fig5} the  
heavier Higgs boson of two heavy Higgs bosons is (dominantly) CP--even 
in the case of $\Phi_{A\mu}=0$ and $\pi/2$, respectively. 
The CP$\tilde{\rm T}$--even observables $\sigma_5$ and $\sigma_{11}$ 
show a sign change at the Higgs--boson resonance pole, while 
the CP$\tilde{\rm T}$--odd observables show a typical peak.

\subsection{The lightest Higgs boson}

According to a detailed analysis \cite{SL1} of Higgs boson decays, the width of
the lightest Higgs boson $H_1$ is of the order of MeV for intermediate
$\tan\beta$, which is smaller than a typical muon energy resolution.
The lightest Higgs boson decays dominantly into $b\bar{b}$ and sub-dominantly
into $\tau^+\tau^-$ with the (almost) fixed branching fractions and
the interference between the Higgs--exchange and the $\gamma$ and $Z$--boson 
exchanges is negligible because of the small $b$ and $\tau$ masses
compared to the Higgs--boson mass, even if the muon beam
polarization is employed.  Furthermore, $\Gamma(H_1\rightarrow 
f_R\bar{f}_R)=\Gamma(H_1\rightarrow f_L\bar{f}_L)$ at the tree level so that
any further information on the CP property of the lightest Higgs boson
will not be obtained by measuring the difference between the final 
$(++)$ and $(--)$ helicity configurations\footnote{The identification of the
$\tau^+\tau^-$ helicities is, however, quite useful to reduce the continuum 
backgrounds from the $\gamma$ and $Z$ exchanges \cite{BHZ}.}.\\

A convolution of the Higgs--boson production cross section with the beam 
energy distribution tends to reduce the production rate as given by the 
expression 
\begin{eqnarray}
\Sigma_{H_1 eff} \approx \frac{\pi\,\Gamma_{H_1}}{2\sqrt{2\pi}\sigma_E}
          \,\Sigma_{H_1}(m_{H_1})\,,
\label{eq:lightest Higgs}
\end{eqnarray}
where $\sigma_E$ is the muon beam energy resolution and the
distribution $\Sigma_{H_1}$ is given by
\begin{eqnarray}
\Sigma_{H_1} (m_{H_1})
   =  \left(1+P_L\bar{P}_L\right)\, C_3 
   + \left( P_L+\bar{P}_L\right)\, C_4 +
 \,P_T\bar{P}_T\left[\,\cos\eta\, C_{15}
   - \sin\eta\, C_{16}\right]\,.
\end{eqnarray}
Every $C_i$ is proportional to the second power of $m_{H_1}/\Gamma_{H_1}$ 
and factored into the production and decay parts on the lightest 
Higgs--boson resonance pole. So, 
apart from the factor depending on
the beam energy  resolution, 
the CP property of the lightest Higgs boson
can be directly investigated through the Higgs--boson resonance production 
by polarized muon and anti--muon collisions as well as through final 
fermion spin--spin correlations with high efficiencies. 
The reduction factor in Eq.~(\ref{eq:lightest Higgs}) clearly 
shows the importance of having a very good beam 
energy resolution in order to obtain 
the spin--0 Higgs--exchange signal events clearly distinguished 
from the spin--1 $\gamma/Z$--exchange background events.
Since several works \cite{SL2,SONI,GGP} along this line have been 
already done, we close this section to referring to the
relevant recent works.

\section{Summary and Conclusion}
\label{sec:conclusion}

We have performed a systematic investigation of the production of a 
third--generation fermion--pair in polarized $\mu^+\mu^-$ 
collisions so as to probe the explicit CP violation in the MSSM Higgs 
sector, induced  radiatively by soft trilinear interactions
related to squarks of the third generation.
We have classified all the observables for 
probing the CP property of the Higgs bosons constructed by the initial 
muon beam polarization along with the final fermions of no polarization
and of equal helicity, respectively. 
The polarization observables have turned out to allow for complete 
determination of the CP property of the Higgs bosons.
Furthermore, we have found that the interference between the 
spin--0 Higgs--boson and spin--1 $\gamma/Z$ contributions  
can provide a powerful and independent means for the 
determination of the CP property of two heavy Higgs boson 
in the top--quark pair production with the c.m. energy near the 
(almost) degenerate heavy Higgs--boson resonances.  
On the other hand, there is no sizable interference between the 
lightest Higgs--boson and spin--1 $\gamma/Z$ contributions so that
the CP property of the lightest Higgs boson 
can be optimally measured on its pole by using the initial muon beam 
polarization or the final fermion spin--spin correlations 
with high efficiencies while keeping a very good beam energy resolution.\\

In conclusion, the third--generation fermion--pair production 
in $\mu^+\mu^-$ collisions equipped with initial muon beam polarization
and final--fermion spin correlations
provides a powerful probe of the CP property of the Higgs bosons
in the MSSM with explicit CP violation.

\section*{Acknowledgements}

S.Y.C wishes to acknowledge financial support of the 1997 Sughak program
of the Korea Research Foundation.  E.A thanks KIAS for the great hospitality 
extended to her while this work was being performed.


\newpage

\addtocounter{figure}{2}
\vskip -1.5cm
\begin{figure}[htb]
 \begin{center}
\hbox to\textwidth{\hss\epsfig{file=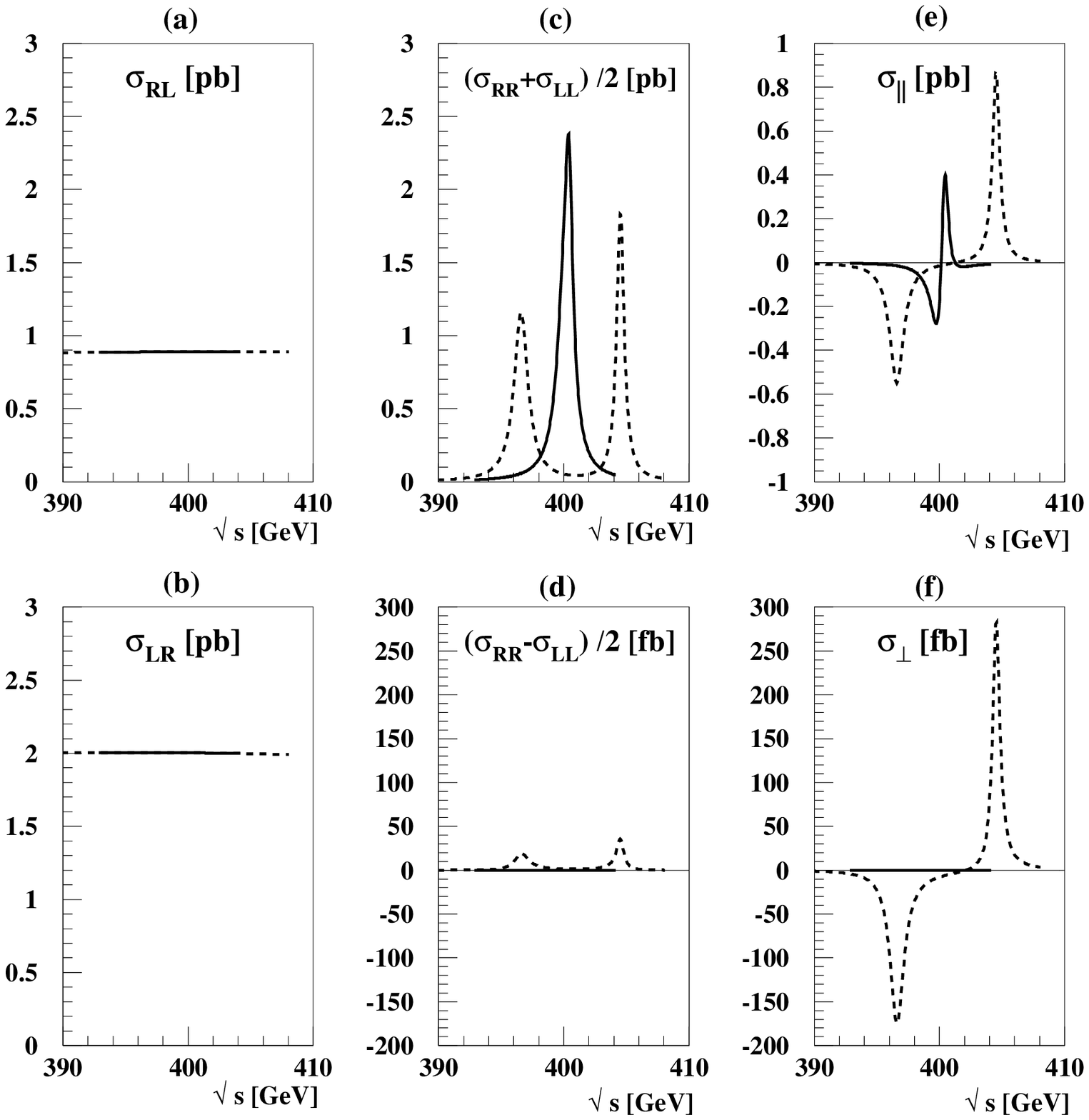,width=16.5cm,height=12cm}\hss}
 \end{center}
 \vskip -1cm
\caption{\it The six independent observables 
          of definite CP and CP$\tilde{T}$ parities; 
	  (a) $\sigma_{_{RL}}[++]$, (b) $\sigma_{_{LR}}[\,++]$,  
          (c) $[\,\sigma_{_{RR}}+\sigma_{_{LL}}]/2[\,++]$,  
          (d) $[\,\sigma_{_{RR}}-\sigma_{_{LL}}]/2[\,--]$,  
          (e) $\sigma_{\Vert}[\,++]$ and  (f) $\sigma_{\bot}[-+]$,
	  near the region of the heavy Higgs--boson 
	  resonances for $\tan\beta=3$ and the SUSY parameter set
	  (\ref{eq:parameter set}). The definition of the observables 
          is given in Eq.~(\ref{eq:observable1}). The solid line
	  is for $\Phi_{A\mu}=0$ and the dashed line for $\Phi_{A\mu}=
	  \pi/2$.}
 \label{fig:fig1}
\end{figure}

\vskip -1cm

\begin{figure}[htb]
 \begin{center}
\hbox to\textwidth{\hss\epsfig{file=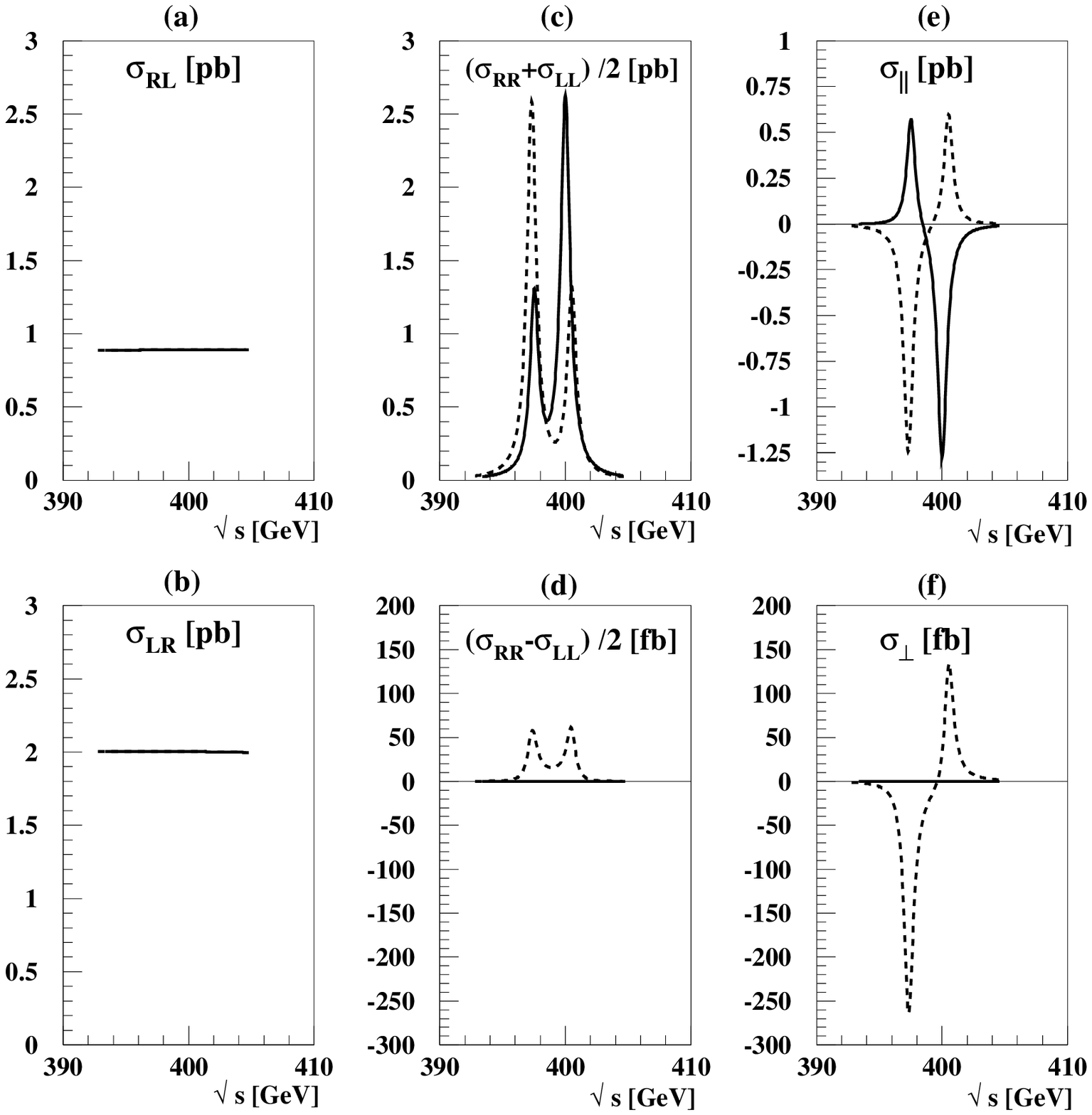,width=16.5cm,height=12cm}\hss}
 \end{center}
 \vskip -1cm
\caption{\it The same six independent observables 
          as in Fig.~\ref{fig:fig1} near the region of the heavy 
	  Higgs--boson resonances for $\tan\beta=10$ and the SUSY parameter 
	  set (\ref{eq:parameter set}).The solid line
	  is for $\Phi_{A\mu}=0$ and the dashed line for $\Phi_{A\mu}=
	  \pi/2$.}
 \label{fig:fig2}
\end{figure}

\vskip -1.5cm
\begin{figure}[htb]
 \begin{center}
\hbox to\textwidth{\hss\epsfig{file=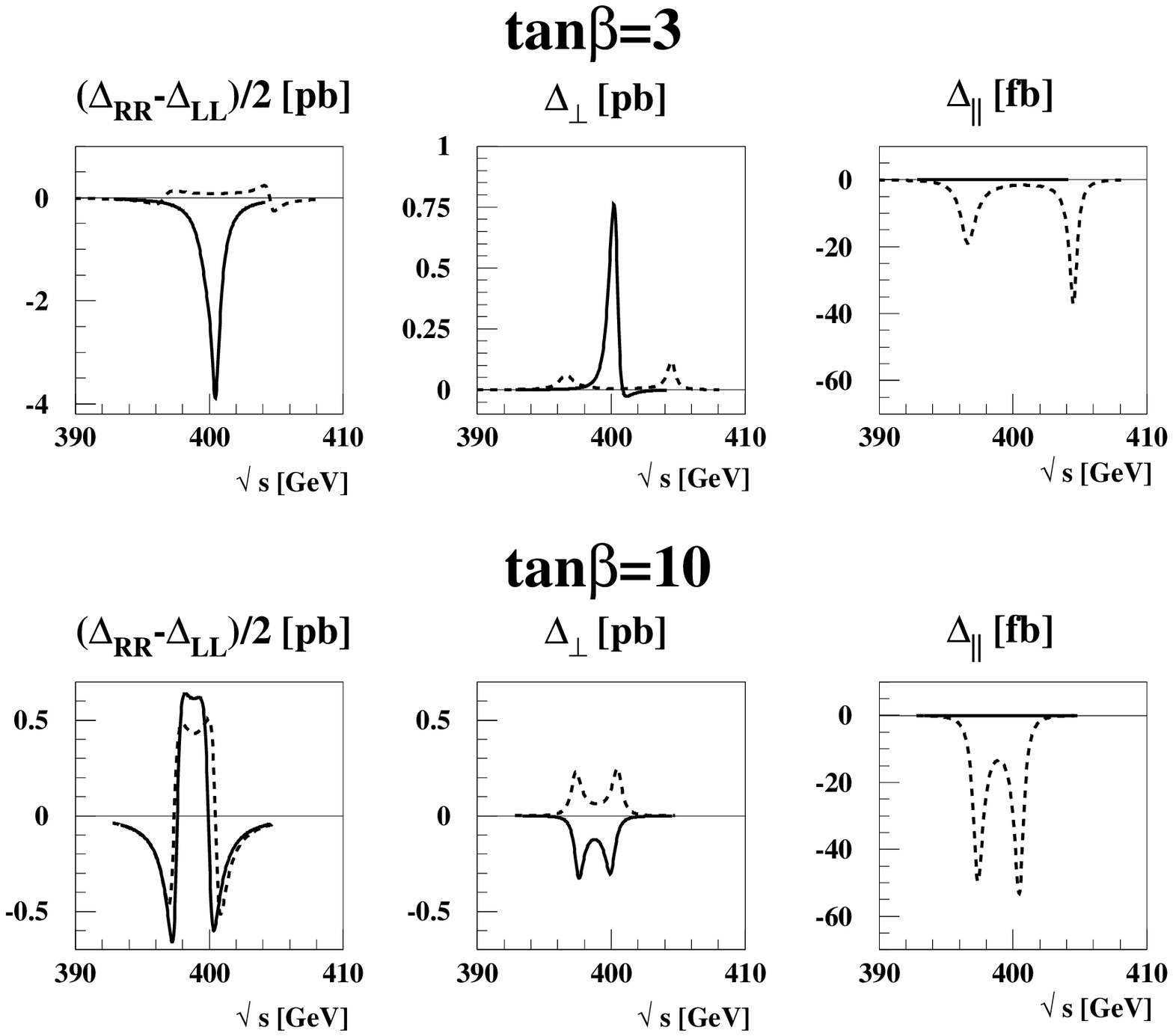,width=16cm,height=13cm}\hss}
 \end{center}
 \vskip -2.5cm
\caption{\it The three observables 
          of definite CP and CP$\tilde{T}$ parities; 
          $[\Delta_{_{RR}}-\Delta_{_{LL}}]/2[\,++]$ (left),  
	  $\Delta_{\bot}[+-]$ (middle), and $\Delta_{\Vert}[--]$ (right) 
	  near the region of the heavy Higgs--boson resonances 
	  for $\tan\beta=3$ (upper set) and $\tan\beta=10$ (lower set)
	  with the SUSY parameter set (\ref{eq:parameter set}). 
	  The definition of the observables is given in 
	  Eq.~(\ref{eq:observable2}). The solid line
	  is for $\Phi_{A\mu}=0$ and the dashed line for $\Phi_{A\mu}=
	  \pi/2$.}
 \label{fig:fig3}
\end{figure}

\vskip 2mm

\vskip -1.5cm
\begin{figure}[htb]
 \begin{center}
\hbox to\textwidth{\hss\epsfig{file=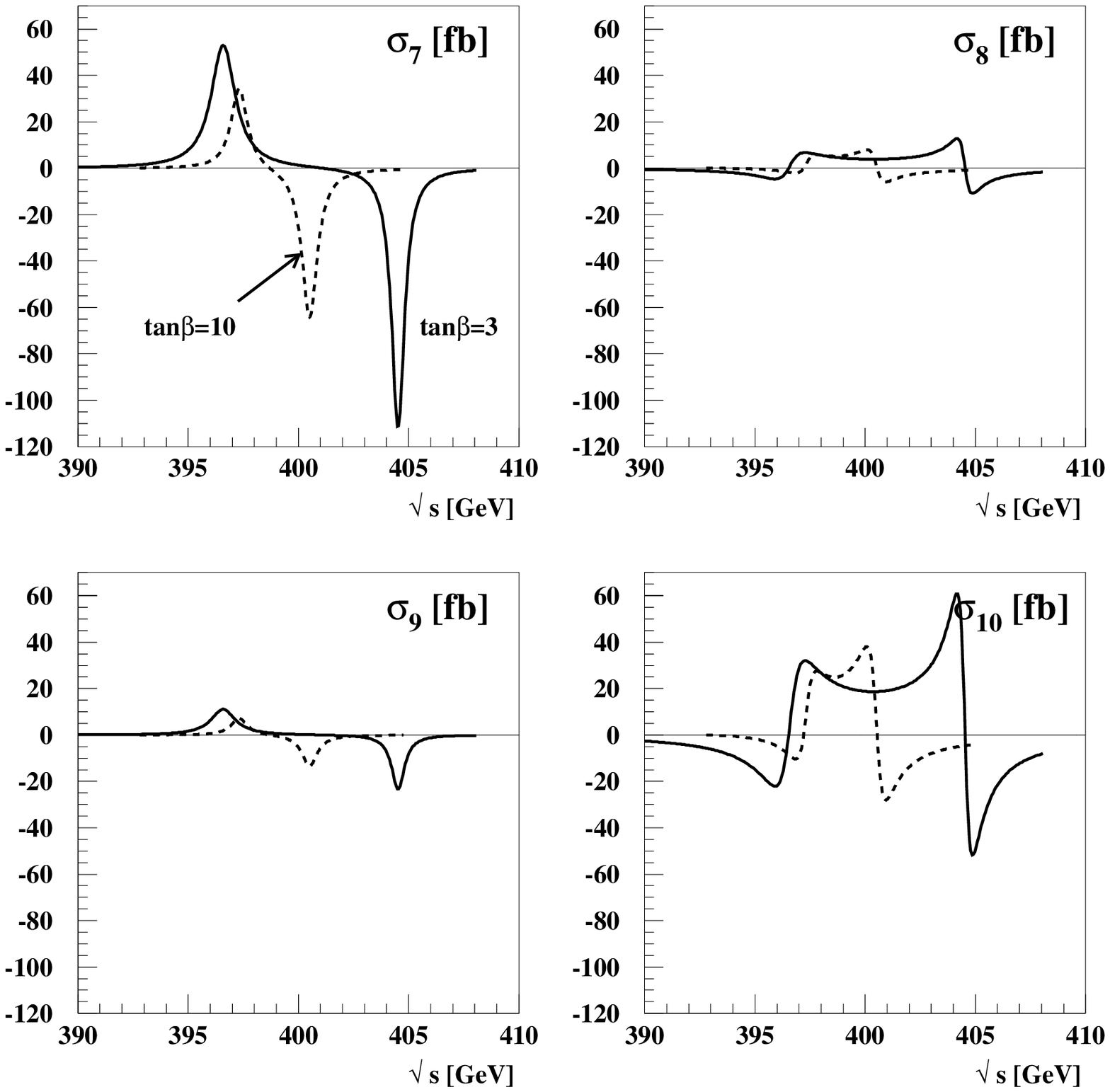,width=16cm,height=13cm}\hss}
 \end{center}
 \vskip -1.5cm
\caption{\it The four CP--odd
          observables $\{\sigma_7[--], \sigma_8[-+], 
	  \sigma_9[--], \sigma_{10}[-+]\}$ 
	  for the $t \bar{t}$ mode near the heavy Higgs--boson resonances 
	  with the SUSY parameter set (\ref{eq:parameter set}). 
	  The solid line is for $\tan\beta=3$  and the dashed line
	  for $\tan\beta=10$. The definition of the observables is 
	  given in Eq.~(\ref{eq:observable3}). The CP phase $\Phi_{A\mu}$
	  is taken to $\pi/2$.}
\label{fig:fig4}
\end{figure}

\vskip 2mm

\begin{figure}[htb]
 \begin{center}
\hbox to\textwidth{\hss\epsfig{file=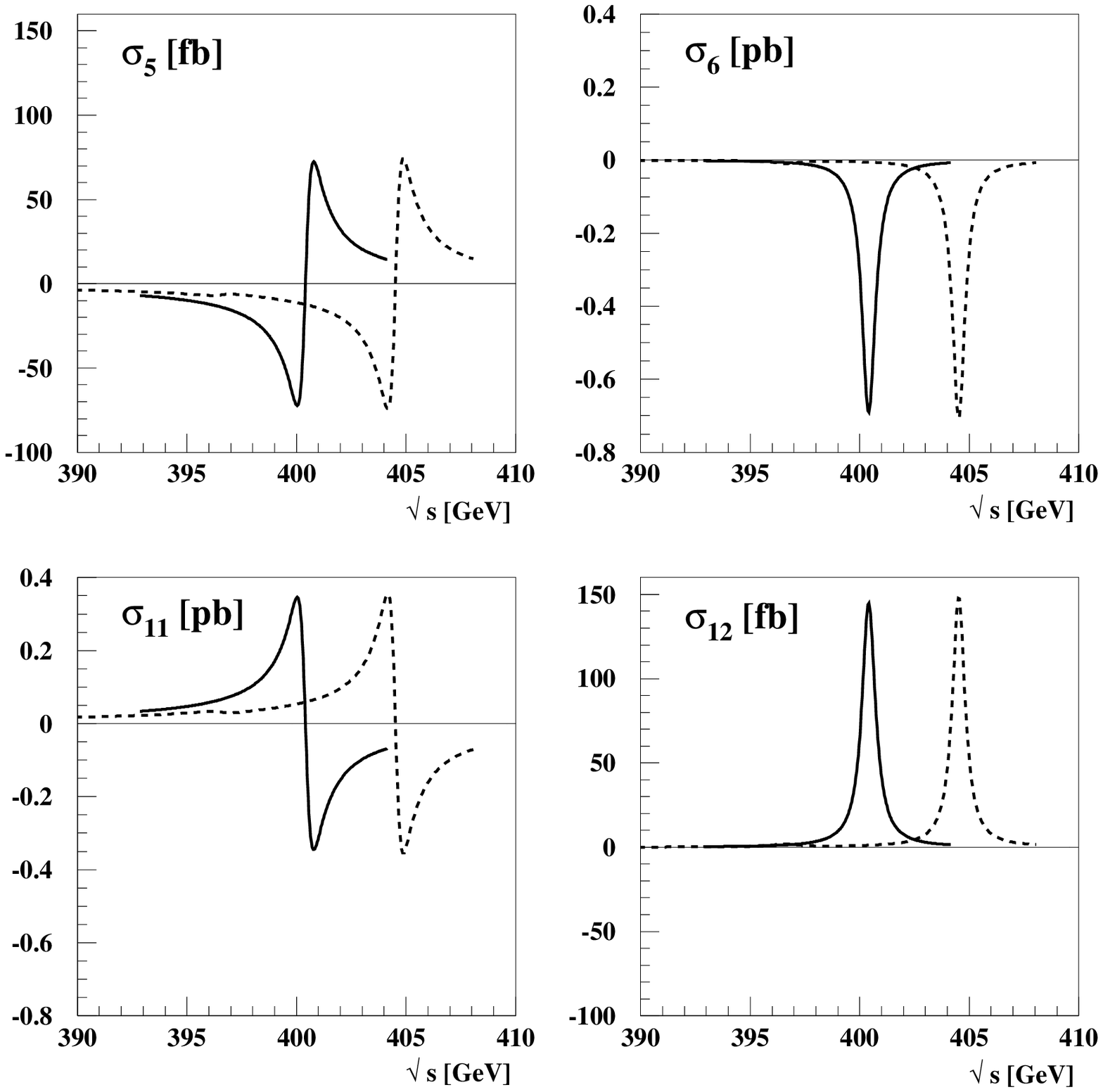,width=16cm,height=13cm}\hss}
 \end{center}
 \vskip -1.5cm
\caption{\it The four CP--even 
          observables $\{\sigma_5[++],\sigma_6[+-],
	  \sigma_{11}[++],\sigma_{12}[+-]\}$ 
	  for the $t \bar{t}$ mode near the heavy Higgs--boson resonances 
	  for $\tan\beta=3$ with the SUSY parameter set 
	  (\ref{eq:parameter set}). 
	  The solid line is for $\Phi_{A\mu}=0$  and the dashed line
	  for $\Phi_{A\mu}=\pi/2$. The definition of the observables is 
	  given in Eq.~(\ref{eq:observable3}). }
\label{fig:fig5}
\end{figure}

\vskip 2mm


\begin{thebibliography}{99}

\bibitem{Hunter} See, J.F. Gunion, H.E. Haber, G. Kane and S. Dawson, 
   {\em `Higgs Hunter's Guide'}, Addison-Wesley Publishing Company, (1990), 
   and references therein; M. Spira and P.M. Zerwas, hep--ph/9803257.   

\bibitem{Kaon} J.H. Christenson, J.W. Cronin, V.L. Fitch and R. Turlay,
   Phys. Rev. Lett. {\bf 13}, 138 (1964); for reviews, see, for example,
   P.K. Kabir, {\it The CP Puzzle}, (Academic Press, London and New York,
   1968); W. Grimus, Fortschr. Phys. {\bf 36}, 201 (1988); E.A. Paschos 
   and U. T\"{ur}ke, Phys. Rep. {\bf 178}, 147 (1989); B. Winstein
   and L. Wolfenstein, Rev. Mod. Phys. {\bf 65}, 1113 (1993);
   G.D. Barr {\it et al.}, NA31 Collaboration, Phys. Lett. {\bf B317}, 233 
   (1993); A. Alavi--Harati {\it et al.}, KTeV Collaboration, Phys. Rev.
   Lett. {\bf 83}, 22 (1999); V. Fanti {\it et al.}, NA48 Collaboration,
   Phys. Lett. {\bf B465}, 335 (1999).

\bibitem{B-meson} For pedagogical introduction to CP violation in the
   B--meson system, see M. Neubert, Int. J. Mod. Phys. {\bf A11}, 4173 (1996);
   A.J. Buras, hep--ph/9806471 and references therein.

\bibitem{EWBG} N. Ruis and V. Sanz, Nucl. Phys. {\bf B570}, 155 (2000)
   and references therein.

\bibitem{EXCP1} A. Pilaftsis, Phys. Lett. {\bf B435}, 88 (1998), 
    hep--ph/9805373; Phys. Rev. {\bf D58}, 096010 (1998), hep--ph/9803297.

\bibitem{EXCP2} A. Pilaftsis and C.E.M. Wagner, Nucl. Phys. {\bf B553}, 
   3 (1999), hep--ph/9803297. 

\bibitem{EXCP3} D.A. Demir, Phys. Rev. {\bf D60}, 055006 (1999), 
   hep--ph/9901389. 

\bibitem{CDL} S.Y. Choi, M. Drees and J.S. Lee, Phys. Lett. {\bf B481},
57 (2000).

\bibitem{EXCP4} M. Carena, J. Ellis, A. Pilaftsis and C.E.M. Wagner, 
   hep--ph/0003180.

\bibitem{CKP} D. Chang, W.--Y. Keung and A. Pilaftsis, Phys. Rev. Lett. 
   {\bf 82}, 900 (1999), erratum: {\bf 83}, 3972 (1999), hep--ph/9811202;
    A. Pilaftsis, Phys. Lett. {\bf B471}, 174 (1999), hep--ph/9909485.

\bibitem{XEDM1} P.~Nath, Phys.~Rev.~Lett.~{\bf 66}, 2565 (1991);
   Y.~Kizukuri and N.~Oshimo, Phys.~Rev.~D {\bf 45}, 1806 
   (1992); {\bf 46}, 3025 (1992).

\bibitem{XEDM2} S.~Dimopoulos and G.F.~Giudice, Phy.~Lett.~B {\bf 357}, 
   573 (1995); A.~Cohen, D.B.~Kaplan and A.E.~Nelson, {\it ibid.}~B 
   {\bf 388}, 599 (1996); A.~Pomarol and D.~Tommasini, 
   Nucl.~Phys.~{\bf B488}, 3 (1996).

\bibitem{XEDM3} T.~Ibrahim and P.~Nath, Phys.~Lett.~B {\bf 418}, 98 (1998);
   Phys.~Rev.~D {\bf 57}, 478 (1998); D {\bf 58}, 019901 (1998);
   T.~Falk and K.A.~Olive, Phys.~Lett.~B {\bf 439}, 71 (1998); 
   M.~Brhlik, G.J.~Good and G.L.~Kane, {\it ibid.}~D {\bf 59}, 
   115004 (1999); S.~Pokorski, J.~Rosiek and C.A.~Savoy, 
   Nucl. Phys. {\bf B570}, 81 (2000).

\bibitem{EXCP_FC} A. Pilaftsis, Phys. Rev. Lett. {\bf 77}, 4996 (1997), 
   hep--ph/9603328; K.S. Babu, C. Kolda, J. March--Russell and F. Wilczek, 
   Phys. Rev. {\bf D59}, 016004 (1999), hep--ph/9804355; S.Y. Choi and 
   M. Drees, Phys. Rev. Lett. {\bf 81}, 5509 (1998), hep--ph/9808377;
   J.F. Gunion and J. Pliszka, Phys. Lett. {\bf B444}, 136 (1998), 
   hep--ph/9809306; C.A. Boe, O.M. Ogreid, P. Osland and J. Zhang, 
   Eur. Phys. J. {\bf C9}, 413 (1999), hep--ph/9811505;
   B. Grzadkowski, J.F. Gunion and J. Kalinowski, Phys. Rev. {\bf D60},
   075011 (1999), hep--ph/9902308.

\bibitem{SL1} S.Y. Choi and J.S. Lee, Phys. Rev. {\bf D61}, 015003 (2000). 
   
\bibitem{SL2} S.Y. Choi and J.S. Lee, Phys. Rev. {\bf D61}, 111702  (2000);
    {\it ibid.} {\bf 61}, 115002 (2000); 
    {\it ibid.} {\bf 62}, 036005 (2000).

\bibitem{DM} A. Dedes and S. Moretti, Phys. Rev. Lett. {\bf 84}, 22 (2000);
Nucl.Phys. {\bf B576}, 29  (2000).

\bibitem{MUCOL} V. Barger, M.S. Berger, J.F. Gunion and T. Han,
   Phys. Rep. {\bf 281}, 1 (1997).

\bibitem{S_H}  V. Barger, M.S. Berger, J.F. Gunion, and T. Han,
   Phys. Rev. Lett. {\bf 75}, 1462 (1995); J.F. Gunion, in 
   {\it Proceedings of the 5th International Conference on Physics
   Beyond the Standard Model, Balholm, Norway, 1997}, edited by
   G. Eigen, P. Osland and B. Stugu (AIP, Woodbury, New York, 1997),
   p. 234.

\bibitem{SONI} D. Atwood and A. Soni, Phys. Rev. D {\bf 52}, 6271 (1995);
   A. Pilaftsis, Phys. Rev. Lett. {\bf 77}, 4996 (1996); Nucl. Phys.
   {\bf B504}, 61 (1997); S.Y. Choi and M. Drees, Phys. Rev. Lett. 
   {\bf 81}, 5509 (1998).

\bibitem{GGP} B. Grzadkowski, J.F. Gunion and  J. Pliszka, 
   Nucl. Phys. {\bf B583}, 49 (2000), hep--ph/0003091; 
   hep--ph/0004034 and references therein.

\bibitem{ERI} E. Asakawa, A. Sugamoto, J. Kamoshita and I. Watanabe,
   Eur. Phys. J. {\bf C14}, 335 (2000), hep--ph/9912373; 
   E. Asakawa, A. Sugamoto and I. Watanabe, hep--ph/0004005.

\bibitem{DS} S. Dimopoulos and D. Sutter, Nucl.~Phys.~{\bf B 452}
   (1995) 496; H. Haber, Proceedings of the 5th International 
   Conference on Supersymmetries in Physics (SUSY'97), May 1997,
   ed. M. Cveti\'{c} and P. Langacker, hep-ph/9709450.

\bibitem{CW} S. Coleman and E. Weinberg, Phys. Rev. {\bf D7}, 1888 (1973);
   Y. Okada, M. Yamaguchi and T. Yanagida, Prog. Theor. Phys.
   {\bf 85}, 1 (1991); Phys. Lett. {\bf B262}, 54 (1991);
   J. Ellis, G. Ridolfi and F. Zwirner, Phys. Lett. {\bf B257}, 83
   (1991); {\bf B262}, 477 (1991). 

\bibitem{HZ} K.~Hagiwara and D.~Zeppenfeld, Nucl. Phys. {\bf B274}, 1
  (1986).

\bibitem{SZ} L.M.~Sehgal and P.M.~Zerwas, Nucl. Phys. {\bf B183}, 417
   (1981).

\bibitem{BHZ} V. Barger, T. Han and C.-G. Zhou, 
   Phys. Lett. {\bf B480}, 140 (2000), hep--ph/0002042.


\end{thebibliography}
\end{document}